\documentclass[a4paper,11pt]{article}
\usepackage{jheppub} % for details on the use of the package, please
\usepackage[T1]{fontenc} % if needed

\title{\boldmath Conserved Charge Fluctuations and Susceptibilities in Strongly Interacting Matter}

\author[a,b]{Shuzhe Shi}
\author[b,c]{Jinfeng Liao}

\affiliation[a]{Physics Department, Tsinghua University, Beijing 100084, China.}
\affiliation[b]{Physics Department and Center for Exploration of Energy and Matter,
Indiana University, 2401 N Milo B. Sampson Lane, Bloomington, IN 47408, USA.}
\affiliation[c]{RIKEN BNL Research Center, Bldg. 510A, Brookhaven National Laboratory, Upton, NY 11973, USA.}

\emailAdd{shisz12@mails.tsinghua.edu.cn}
\emailAdd{liaoji@indiana.edu}

\usepackage{graphicx}
\usepackage{bm}
\usepackage{color}
\usepackage{amsmath}
\usepackage{epstopdf}
\usepackage{epsfig}

\newcommand{\be}{\begin{eqnarray}}
\newcommand{\ee}{\end{eqnarray}}

 \newcommand{\gsim}{\mathrel{\hbox{\rlap{\lower.55ex \hbox {$\sim$}}
                   \kern-.3em \raise.4ex \hbox{$>$}}}}
\newcommand{\lsim}{\mathrel{\hbox{\rlap{\lower.55ex \hbox {$\sim$}}
                   \kern-.3em \raise.4ex \hbox{$<$}}}}

\newcommand{\ba}{\begin{eqnarray}}
\newcommand{\ea}{\end{eqnarray}}

\usepackage{amssymb}

\abstract
{We study the conserved charge fluctuations, as quantified by the corresponding susceptibilities, in strongly interacting matter as motived by the quark-gluon plasma. Using the gauge-gravity correspondence approach, we study the patterns of conserved charge fluctuations in two types of holographic models for QCD, the D4/D8 and the D3/D7 models. We compute and compare the quark number susceptibilities in both models and find an interesting common feature of the two: at very strong coupling higher order susceptibilities are suppressed and the conserved charge fluctuations become purely Guassian.   
In light of the state-of-the-art  lattice QCD results we also discuss  what we can learn from these susceptibilities about the underlying degrees of freedom in the $1\sim 2T_c$ quark-gluon plasma and examine the viability of different ideas such as holography, quasi-particles, as well as bound states. From analyzes of second order susceptibilities we conclude that the bound states exist and are important in the $1\sim 2T_c$ region. We further construct and make predictions for several ratios of fourth-order susceptibilities that can sensitively reveal such bound states.  }

\begin{document} 
\maketitle
\flushbottom

\section{Introduction}

The many body systems in the strong interaction sector of the Standard Model, described by the Quantum Chromodynamics (QCD), show very rich phase structures and provide fascinating examples of strongly interacting matter \cite{BraunMunzinger:2009zz,Stephanov:2007fk,Itoh:1970uw}. Amongst others a color-deconfined and chiral-symmetric phase at high enough temperature (and low baryonic density), known as the quark-gluon plasma (QGP), is predicted from QCD and has been created and probed in relativistic heavy ion collision experiments at the Relativistic Heavy Ion Collider (RHIC) and the Large Hadron Collider (LHC). Both the first-principle lattice QCD simulations of the QGP and the experimental measurements of its properties have consistently shown that the QGP in the temperature region $T \sim$ a few $T_c$ (the transition temperature $\sim 160\rm MeV$ in QCD~\cite{Aoki:2006br,Bazavov:2011nk}) is still a very strongly interacting system rather than an asymptotically free plasma \cite{sQGP_review,sQGP_ES} and may be dominated by emergent degrees of freedom such as the chromo-magnetic monopoles \cite{Liao:2006ry,Ratti:2008jz,Simic:2010sv}. While the transition to QGP at high T is a crossover, it is expected that there will be a first order phase transition at high baryonic density and low temperature together with a critical end point (CEP) at certain critical temperature and density that separates the crossover region from the first order transition. There are intensive efforts both from lattice QCD and from Beam Energy Scan (BES) experiments at RHIC looking for the critical end point on the QCD phase diagram \cite{Fodor:2004nz,Borsanyi:2012cr,Bazavov:2011nk,deForcrand:2008vr,Pisarski:1983ms,Stephanov:1998dy,Stephanov:1999zu,STAR_BES}. To study the strongly interacting QCD matter in these relevant temperature and density regions is theoretically challenging but of great interest and importance. 

A very useful class of observables for studying such strongly interacting matter includes quantities describing the fluctuations and correlations of conserved charges of the underlying microscopic theories. In the context of QCD (counting the three light flavors), the conserved charges include the baryon number $B$, isospin $I$, strangeness $S$, as well as the electric charge $Q$. All these charges are carried by the quarks (the fundamental fermions) or their various combinations (e.g. mesons/baryons as bound states of quarks), and  the patterns of their fluctuations/correlations  provide sensitive probes to the actual degrees of freedom that carry such charges \cite{Liao:2005pa,Koch:2005vg}. Such patterns are particularly important imprints of potentially emergent degrees of freedom when the system is strongly interacting: a famous example from condensed matter physics is the fractional quantum hall state with emergent quasi-particles carrying fractional electron charges that were experimentally first identified through current fluctuation measurements.    For simplicity we focus in this paper mostly on the baryon number fluctuations which is also simply related to the quark number fluctuations as each quark/anti-quark carries $\pm 1/N_c$ baryonic number and the corresponding baryonic chemical potential and quark chemical potential are related as $\mu_B= N_c \, \mu_q$. Let us consider the thermodynamics of QCD-like matter with $N_f$ flavors of fundamental quarks carrying the baryonic charge, described by the pressure $P(T,\mu)$ as a function of temperature $T$ and quark chemical potential $\mu$. These fluctuations can be quantified by the susceptibilities, defined through the Taylor expansion coefficients of pressure over the chemical potential \cite{Allton:2005gk,Cheng:2008zh,Gavai:2005sd,Borsanyi:2011sw,Bazavov:2012jq}:
\begin{eqnarray}
X^q_n(T,\mu)=\frac{\partial^n(P/T^4)}{\partial (\mu/T)^n}  
\end{eqnarray}
The above dimensionless susceptibilities are closely related to charge densities fluctuations and correlations: $X^q_1 \sim <B>$ (measuring density itself), $X^q_2 \sim <B^2>$ (measuring quadratic fluctuations), etc. When there are multiple charges, the susceptibilities can be easily generalized to all charges, and there will also be cross-charge (off-diagonal) susceptibilities that describe the density correlations, e.g. $X_{B\, S} = \frac{\partial^2 (P/T^4)}{\partial (\mu_B/T)\partial(\mu_S/T)} \sim <B\, S>$.  Of particular interest is the zero density limit of these susceptibilities which can be directly evaluated by lattice QCD simulations, and we also introduce the following zero density susceptibilities (per quark flavor for convenient comparison with lattice)
\begin{eqnarray}
\chi^q_n (T) \equiv \frac{X^q_n(T,\mu=0)}{N_f}
\end{eqnarray}
Note in the above by symmetry all the odd susceptibilities vanish at $\mu \to 0$ and only $n=2,4,6,..$ terms survive. 
These susceptibilities are also closely connected with many observables in the heavy ion collision experiments, such as the charge fluctuations and correlations measurements, the search for critical end point, and the freeze-out conditions, etc \cite{Jeon:2003gk,Koch:2008ia,Bzdak:2012ia,STAR_BES,Bazavov:2012vg,Gupta:2011wh}. Let us just mention one such example: the cumulants of net baryonic charge fluctuations proposed for the search of CEP. Let $\sigma$ be the variance and $\kappa$ the kurtosis of the event-by-event net baryon number distribution, then for a thermal system one has the following relation
\begin{eqnarray}
\kappa \sigma^2 = \frac{X^q_4(T,\mu)}{X^q_2(T,\mu)} 
\end{eqnarray}  
It is therefore also interesting to study such ratios of susceptibilities. 

One significant challenge to study these susceptibilities for the quark-gluon plasma in the physically interesting regime, as already mentioned, is that the system is strongly interacting. Indeed the state-of-the-art lattice QCD results \cite{Borsanyi:2011sw,Bazavov:2012jq} show rather distinctive patterns in the $1\sim 2 T_c$ region that can not be understood via two often-used simple benchmarks, namely the free hadronic resonance gases at low temperature and the Stefan-Boltzmann gas of quarks and gluons at asymptotically high temperature (see e.g. discussions in \cite{Kim:2009uu}). To this end, the holography provides a very useful approach for understanding the behaviors of the conserved charge fluctuations in QCD-like strongly interacting matter. As is well known, the holographic models in the past have offered many insights for understanding the properties of the quark-gluon plasma (see e.g. the comprehensive review \cite{CasalderreySolana:2011us} and references therein). There were though only very few studies of these susceptibilities in holographic models using one or another setup \cite{Kim:2009uu,Kim:2010zg,CasalderreySolana:2012np}. In the main part of this paper, we will use the gauge-gravity correspondence approach to study the patterns of conserved charge fluctuations in two major types of holographic models for QCD, the D4/D8 and the D3/D7 models in the Section II and III respectively. We will compute and compare the quark number susceptibilities at various temperature and density in both models and also study the related quantities such as the ratios of them. We will analyze and compare the asymptotic behavior at large coupling and large temperature in both models  and find interesting common features and differences. Another part of the motivation for this work, is to closely examine the implications of the state-of-the-art lattice QCD results for the susceptibilities, which will be done in the Section IV. We will evaluate the viability of different ideas such as holography, quasi-particles, as well as bound states, and discuss what we can learn from them about the underlying degrees of freedom in the $1\sim 2T_c$ quark-gluon plasma.

\section{Susceptibilities from holography: D4/D8 model}

In this section we will evaluate the susceptibilities from one holographic model for QCD based on the D4/D8 branes configuration with one compactified dimension, known as the Sakai-Sugimoto model \cite{Sakai}. In this model, the flavor dynamics is given by $N_f$ $D8-\bar{D8}$ flavor branes in the background fields (the gravity, dilation, and Ramond-Ramond fields) generated by $N_c$ $D4$ branes in the ``probe'' limit $N_f << N_c$. The Sakai-Sugimoto model at zero temperature has been shown to reproduce many results in qualitative or semi-quantitative agreement with QCD \cite{Sakai}. The model has also been extended to finite temperature \cite{FiniteT} and finite baryonic density \cite{FiniteD}, with its phase diagram qualitatively mimicking that in QCD. We refer the readers to the above references for detailed descriptions of the model.  

\begin{figure}[!h]
\begin{center}
\includegraphics[width=0.4\textwidth]{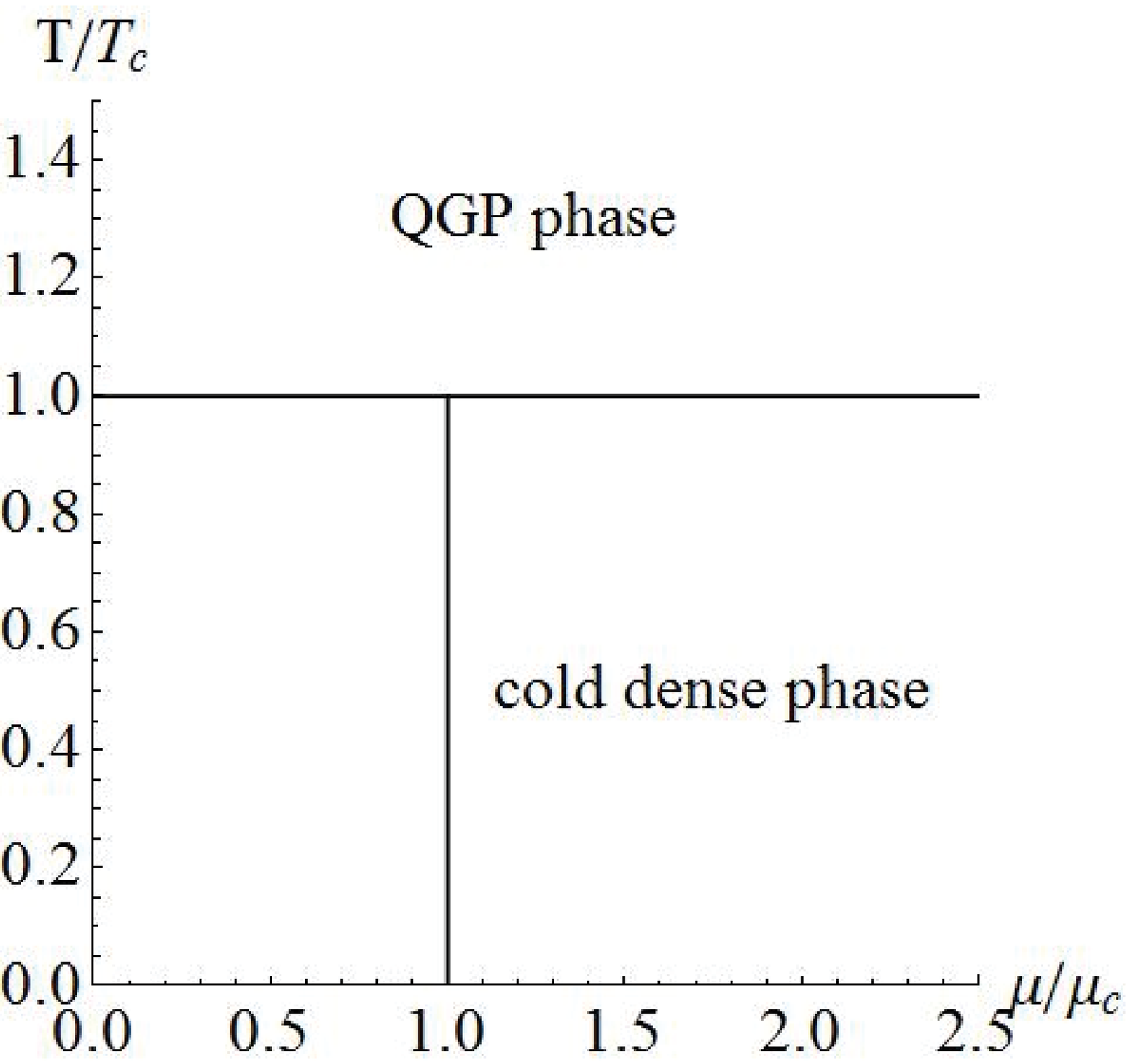}\;
\includegraphics[width=0.4\textwidth]{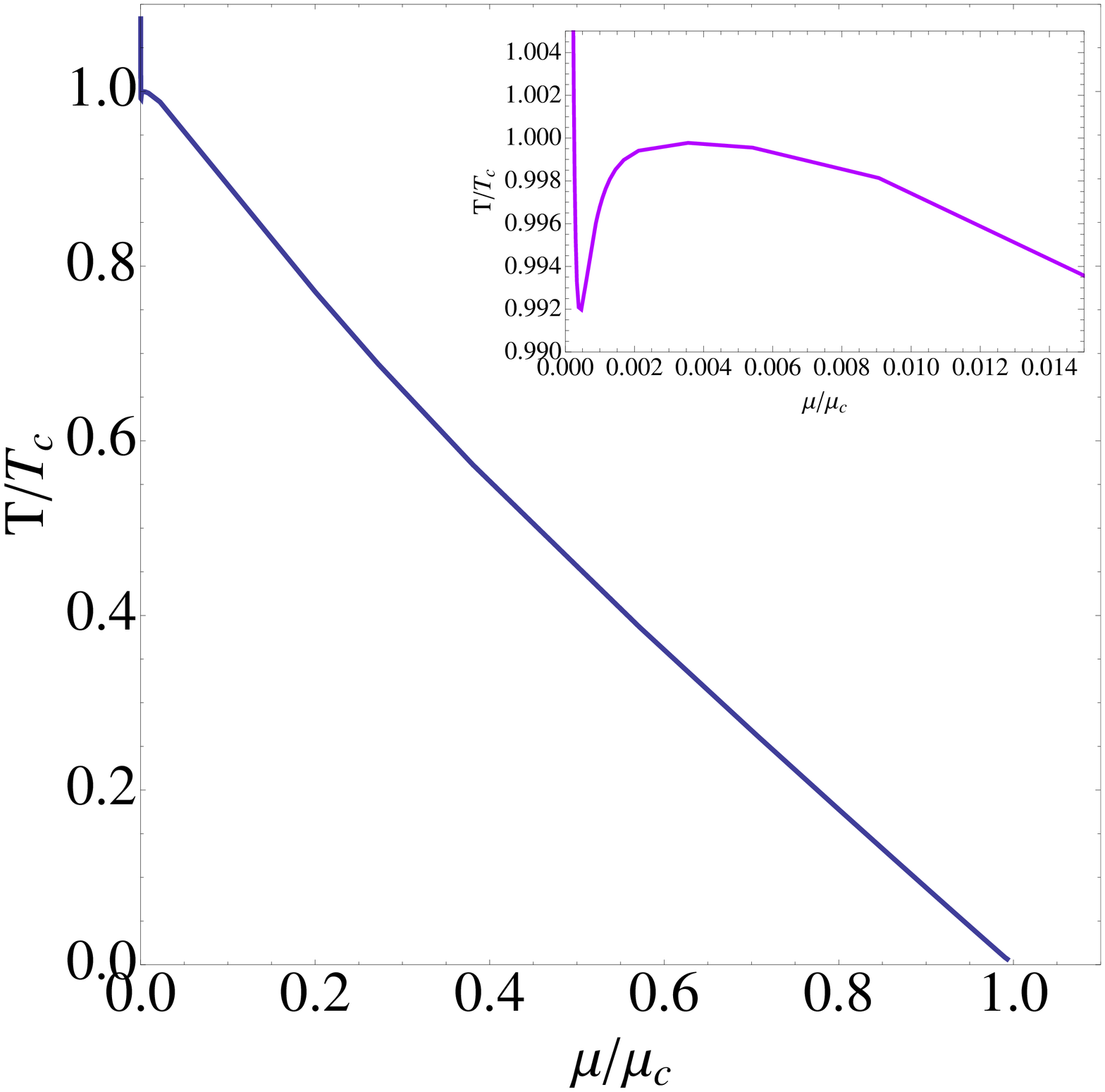}
\caption{The phase diagrams of the D4/D8 model (left) and the D3/D7 model (right, with the inset plot at the top right corner showing the details close to $T_c$). }
\label{fig:phase}
\end{center}
\end{figure}

Let us now discuss the different phases at finite $T$ and $\mu$ in the Sakai-Sugimoto model. At finite temperature there are two possible background geometries (the zero temperature one and the black hole one) and two important  scales involved, the Kaluza-Klein mass scale $M_{KK}$ (from the compactified dimension) and the temperature scale $T$ (ultimately associated with the black hole horizon). At low temperature the zero temperature geometry has the smallest action and is thermodynamically preferred: this is the confined phase. When temperature is high enough $T > T_c \equiv M_{KK}/(2\pi)$ the black hole geometry will ``win'' and the system is in the deconfined phase. By turning on a gauge field in the bulk one induces global conserved (baryonic) charge density (and the related chemical potential) in the boundary field theory and mimics the finite density. This does not change the confined/deconfined phase distinction determined by temperature alone. In the confined phase, however, when the chemical potential is high enough $\mu>\mu_c\equiv \lambda M_{KK} / 27\pi$ (with $\lambda=g^2N_c$ the 't Hooft coupling), there is a new phase (the cold dense phase) that has nonzero charge density and resembles the proposed ``quarkyonic phase'' \cite{McLerran:2007qj}. In the {\em ``vacuum phase''} $T<T_c$ and $\mu<\mu_c$ thermodynamic quantities are independent of $T,\mu$ and therefore all susceptibilities vanish. We will focus on the {\em ``quark-gluon plasma (QGP) phase''} at $T>T_c$ and the {\em ``cold dense phase''} at $T<T_c$ and $\mu>\mu_c$ and obtain the susceptibilities in each phase respectively. The phase diagram is demonstrated in Fig.\ref{fig:phase} (left panel).     

For convenience and simplicity of the presentation, we hereby define some dimensionless variables as following:
\begin{eqnarray}
\widetilde T = \frac{T}{T_c} =  \frac{T}{M_{KK}/2\pi}= \frac{4\pi}{3} \frac{T R}{2M_{KK} R/3} \quad , \quad  \widetilde u_T = \frac{U_T}{(2M_{KK} R/3)^2 R} = \widetilde T^2 \quad ,
\end{eqnarray}
\begin{eqnarray}
\widetilde \mu = \frac{\mu}{\mu_c}=\frac{3 \mu}{(2M_{KK} R/3)^2}\frac{4\pi M_{KK}R^2}{\lambda} \quad , \quad  
\widetilde d = \frac{d}{(2M_{KK} R/3)^5} \quad ,
\end{eqnarray}
and 
\begin{eqnarray}
\widetilde P = \frac{P}{(2M_{KK} R/3)^7} \frac{2^7 3\pi^5M_{KK}^3R^7}{ N_f N_c \lambda^3} \quad .
\end{eqnarray}
In the above, the $P$ is pressure, and $d$ is the quark number density at a given chemical potential $\mu$ (--- noting that the baryon number density would then be $\rho_B=d/N_c$ and the baryonic chemical potential $\mu_B=N_c \mu$).  In addition there are two parameters $R=(\pi g_s N_c l_s^3)^{1/3}$ with $g_s,l_s$ being string coupling and length, and $U_T=(4\pi/3)^2R^3T^2$ the horizon radial coordinate in the black hole geometry. 
More detailed definitions and computations in the Sakai-Sugimoto model specifically related to our discussions can be found in e.g. \cite{Kim:2009uu}. It should be pointed out that for the D4/D8 model we will closely follow the approach of \cite{Kim:2009uu} in which results for susceptibilities at $\mu=0$ in QGP phase and at $\mu=\mu_c$ in cold dense phase  were reported. Here we will compute the susceptibilities generally for any given $T$ and $\mu$.  

\subsection{Susceptibilities in the QGP phase}

For the QGP phase of the D4/D8 model \cite{FiniteT,FiniteD}, the pressure has been computed  at given  temperature $T$ and density $d$ to be (with $_2 F_1$ the hypergeometric function): 
\begin{eqnarray}
\widetilde P (\widetilde T, \widetilde d) = \frac{2}{7}\Big[\frac{2}{3}\frac{\widetilde d^2}{\widetilde u_T^{3/2}}~_2F_1\Big(\frac{3}{10};\frac{1}{2};\frac{13}{10};-\frac{\widetilde d^2}{\widetilde u_T^5} \Big)+ \widetilde u_T\sqrt{\widetilde d^2+\widetilde u_T^5} \Big]
\end{eqnarray}
where the density $d(\widetilde T, \widetilde \mu)$ can be determined from the following relation: 
\begin{eqnarray}
\widetilde \mu = \frac{2\widetilde d}{\widetilde u_T^{3/2}}~_2F_1\Big(\frac{3}{10};\frac{1}{2};\frac{13}{10};-\frac{\widetilde d^2}{\widetilde u_T^5} \Big)
\end{eqnarray}
With the chain rule, $\frac{\partial\widetilde P}{\partial \widetilde\mu}=\frac{\partial \widetilde P/\partial \widetilde d}{\partial\widetilde \mu/\partial \widetilde d}$ and generally $\frac{\partial^{n+1}\widetilde P}{\partial \widetilde\mu^{n+1}}=\frac{\partial (\partial^n\widetilde P/\partial \widetilde\mu^n)/\partial \widetilde d}{\partial\widetilde \mu/\partial \widetilde d}$, we can compute the susceptibilities to arbitrary order defined as:  
\begin{eqnarray} \label{D4D8_Xn}
X^q_n(T,\mu)=\frac{\partial^n(P/T^4)}{\partial (\mu/T)^n}= \left ( 3^{3n-8} 2^{4-n}  \pi^{-1} \right) \, N_c N_f  \lambda^{3-n} \, \widetilde T^{n-4}\frac{\partial^n\widetilde P}{\partial \widetilde\mu^n}
\end{eqnarray}
%%%%%%%%%%%%%
%%%%%%%%%%%%%
A few explicit examples, for the   first to fourth order, are given here:
\begin{eqnarray}
\frac{\partial\widetilde P}{\partial \widetilde\mu}&=&\frac{\widetilde d}{3},\\
\frac{\partial^2\widetilde P}{\partial \widetilde\mu^2}&=&\frac{\widetilde u_T^{3/2}}{6} \, \frac{1}{~_2F_1\Big(\frac{3}{10};\frac{3}{2};\frac{13}{10};-\frac{\widetilde d^2}{\widetilde u_T^5} \Big)},\\
\frac{\partial^3\widetilde P}{\partial \widetilde\mu^3}&=&\frac{9\widetilde d}{52\widetilde u_T^{2}} \,
\frac{~_2F_1\Big(\frac{13}{10};\frac{5}{2};\frac{23}{10};-\frac{\widetilde d^2}{\widetilde u_T^5} \Big)}{~_2F_1\Big(\frac{3}{10};\frac{3}{2};\frac{13}{10};-\frac{\widetilde d^2}{\widetilde u_T^5} \Big)^3}\\
\frac{\partial^4\widetilde P}{\partial \widetilde\mu^4}&=&\frac{3 \widetilde u_T^{-1/2}  }{2^3 13} \,  \frac{13\big(9+\big(1+\frac{\widetilde d^2}{\widetilde u_T^5}\big)^3 ~_2F_1\Big(\frac{3}{10};\frac{3}{2};\frac{13}{10};-\frac{\widetilde d^2}{\widetilde u_T^5} \Big)^2+5\big(\frac{\widetilde d^2}{\widetilde u_T^5}-2\big)~_2F_1\Big(1;-\frac{1}{5};\frac{13}{10};-\frac{\widetilde d^2}{\widetilde u_T^5} \Big)\big)}{75\frac{\widetilde d^2}{\widetilde u_T^5} \big(1+\frac{\widetilde d^2}{\widetilde u_T^5}\big)^3~_2F_1\Big(\frac{3}{10};\frac{3}{2};\frac{13}{10};-\frac{\widetilde d^2}{\widetilde u_T^5} \Big)^5}
\end{eqnarray}

%%%%%%%%%%%%%%%%%%
%%%%%%%%%%%%%%%%%%
\begin{figure}[!h]
\begin{center}
\includegraphics[width=0.31\textwidth]{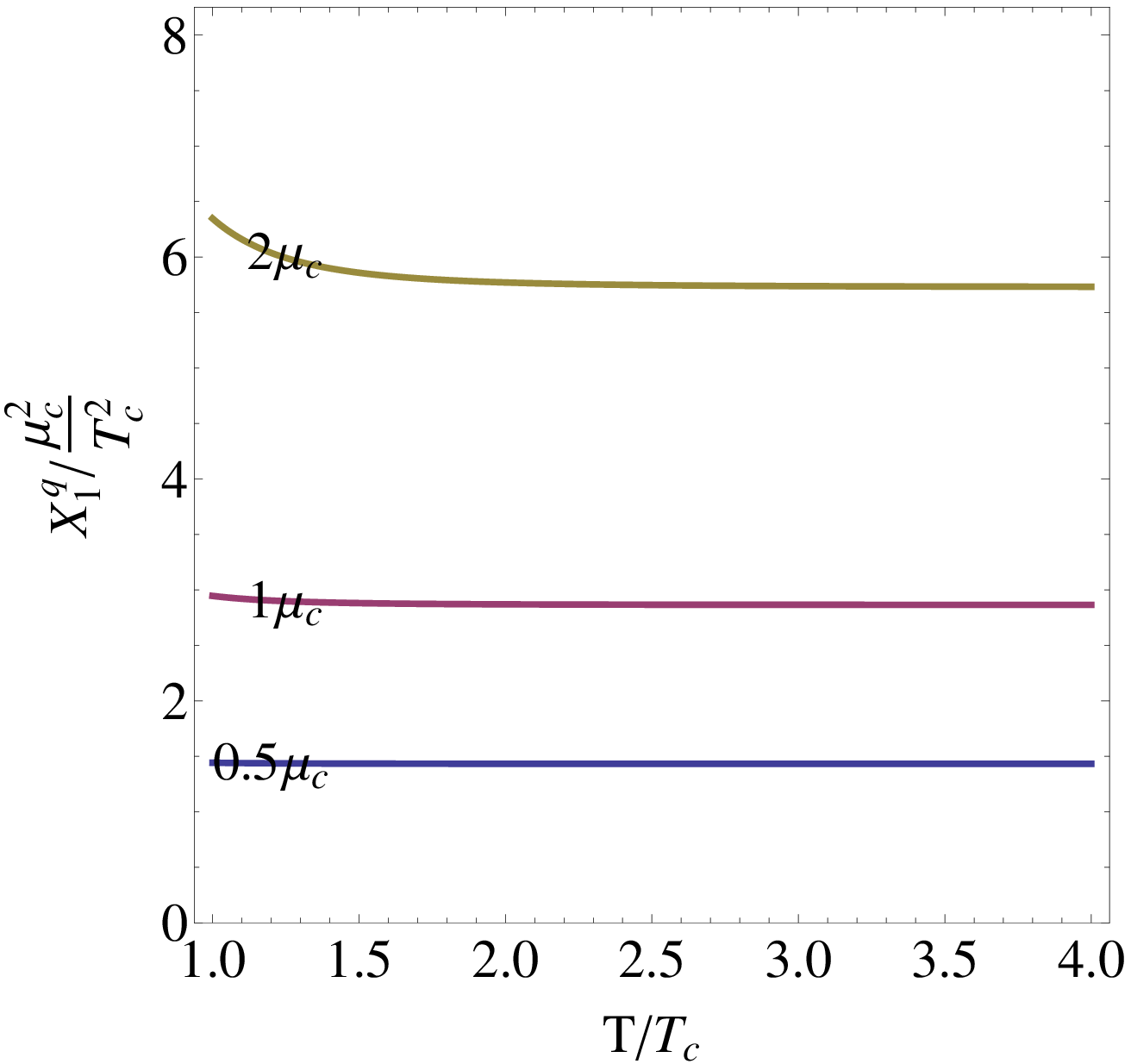} \; 
\includegraphics[width=0.31\textwidth]{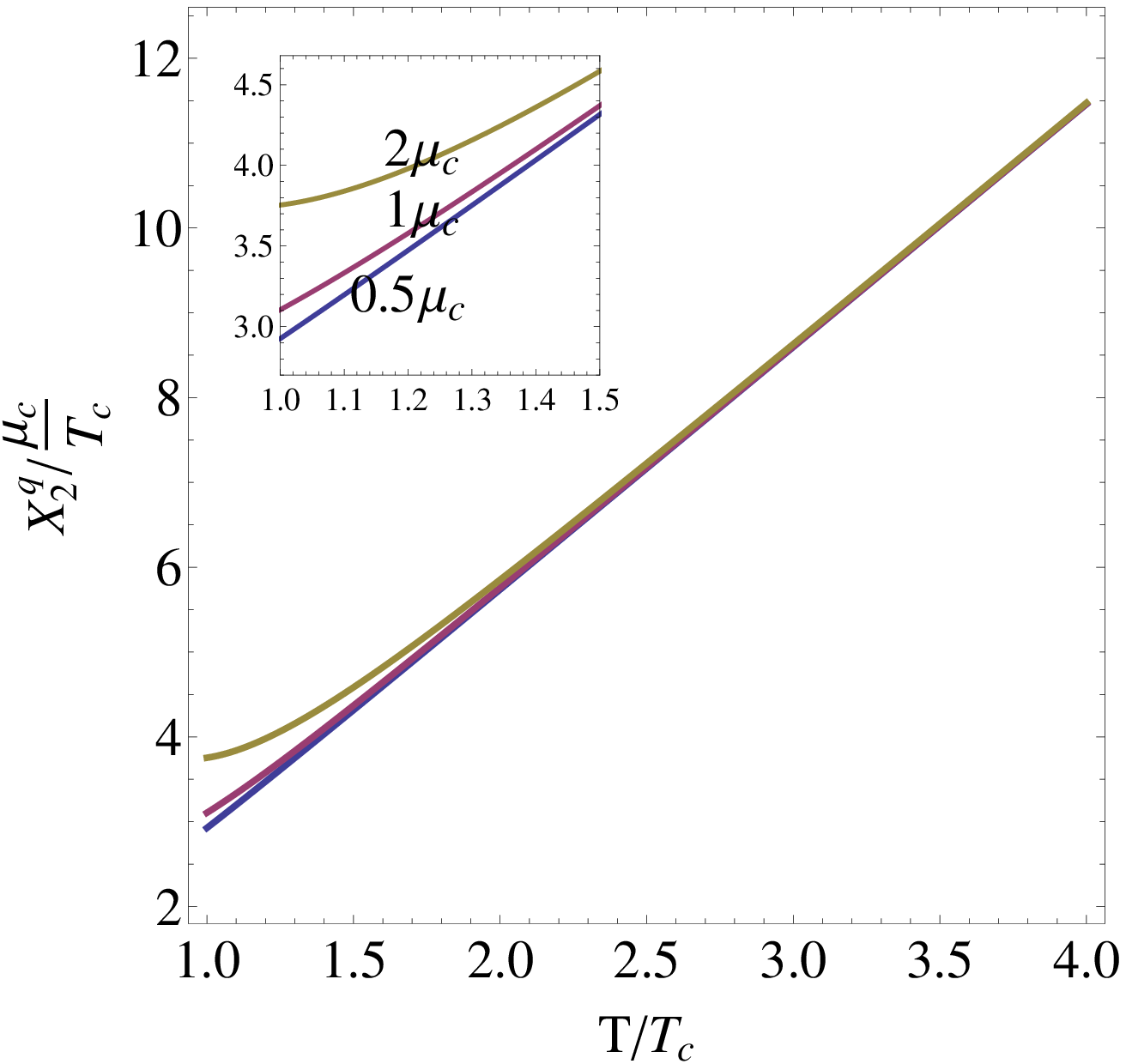} \; 
\includegraphics[width=0.30\textwidth]{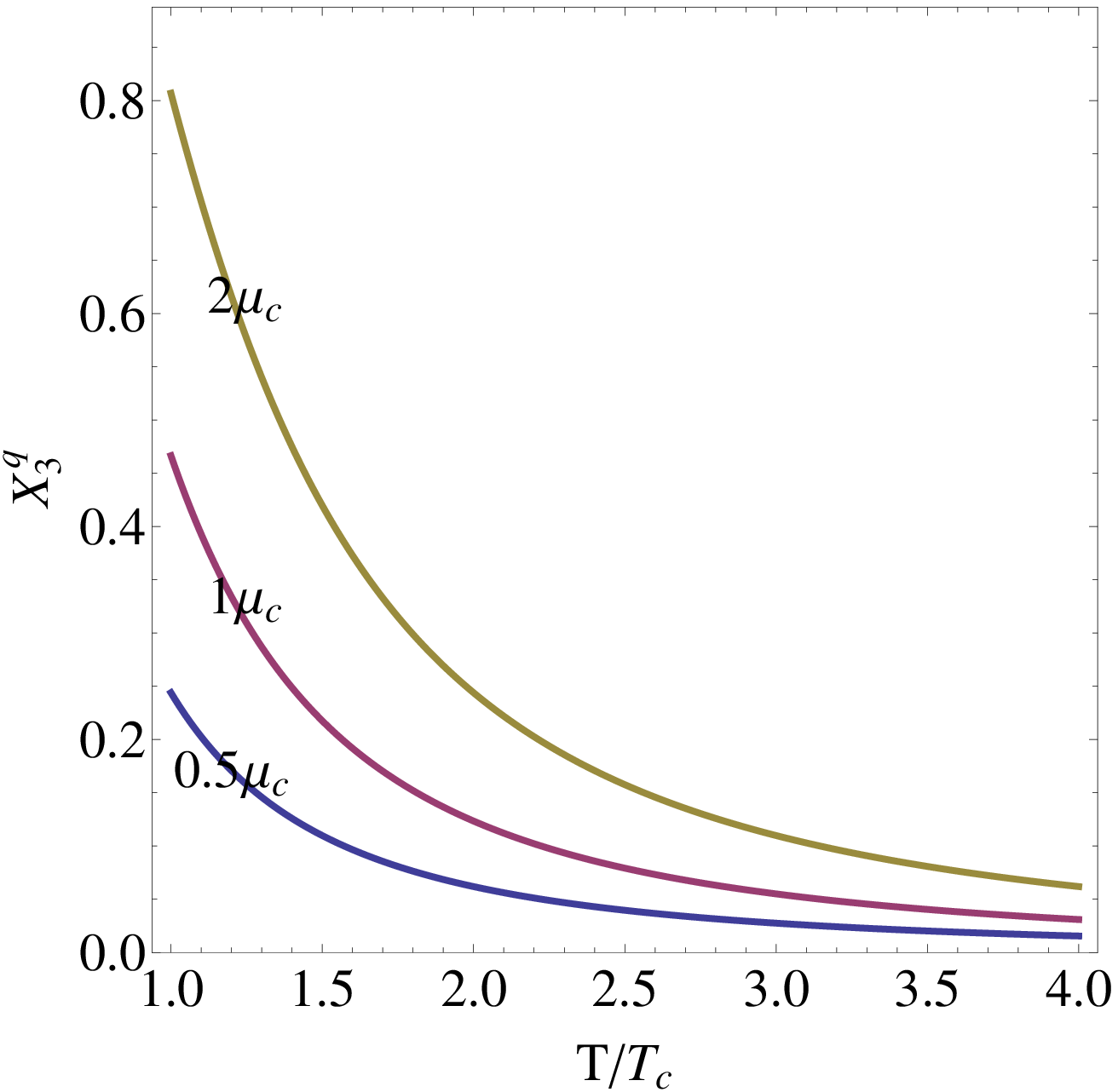} \\
\includegraphics[width=0.30\textwidth]{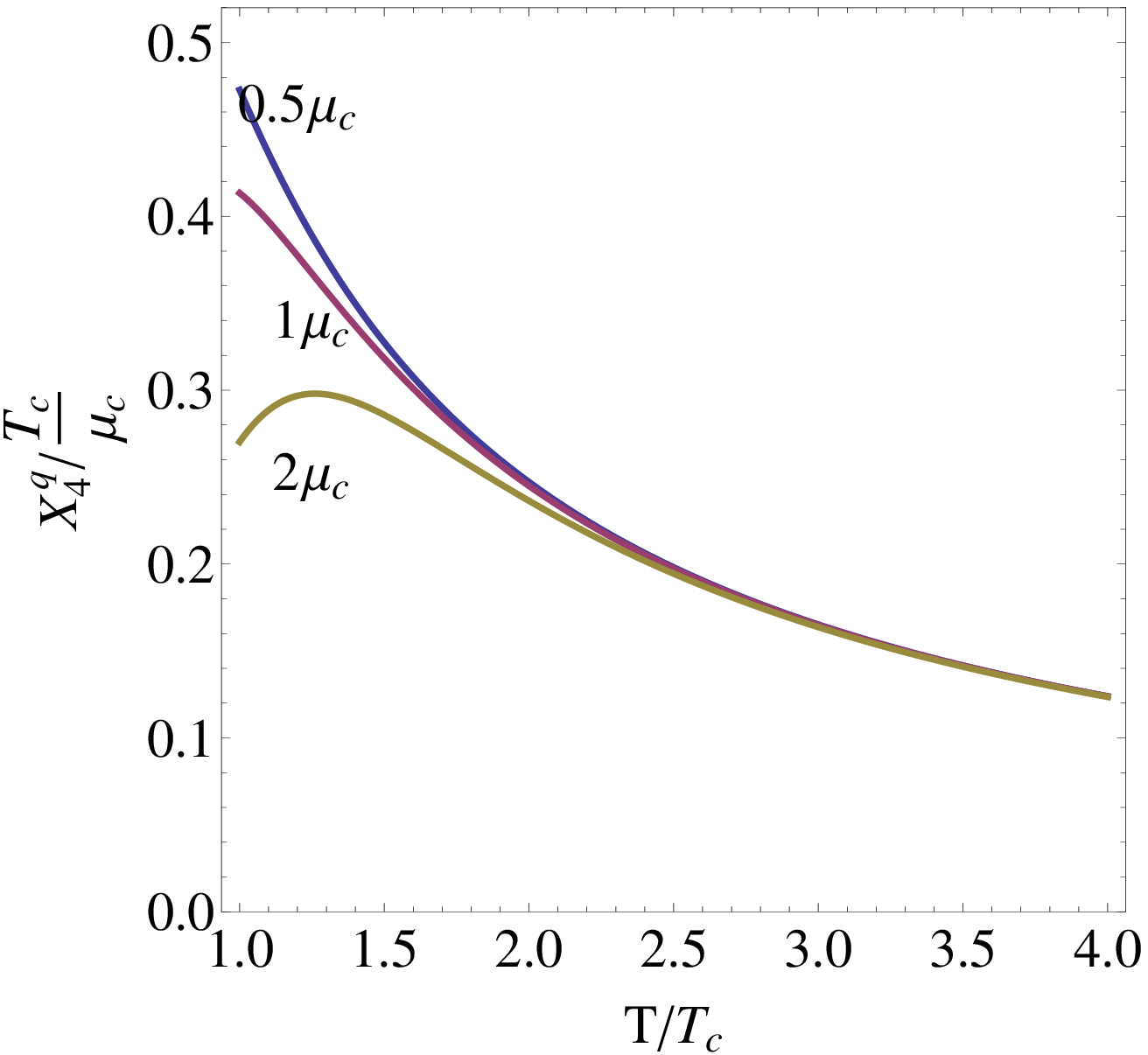} \; 
\includegraphics[width=0.32\textwidth]{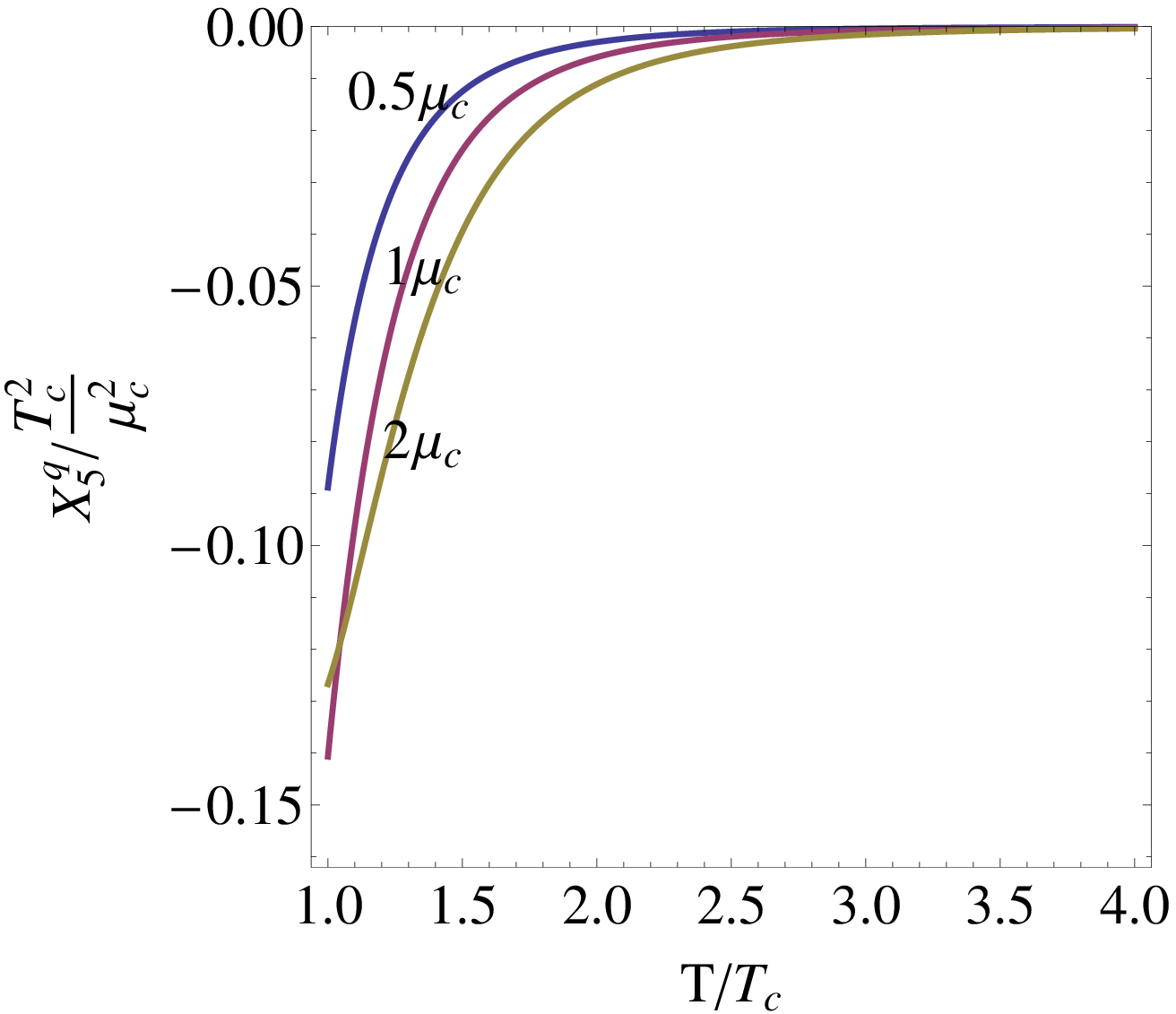} \; 
\includegraphics[width=0.32\textwidth]{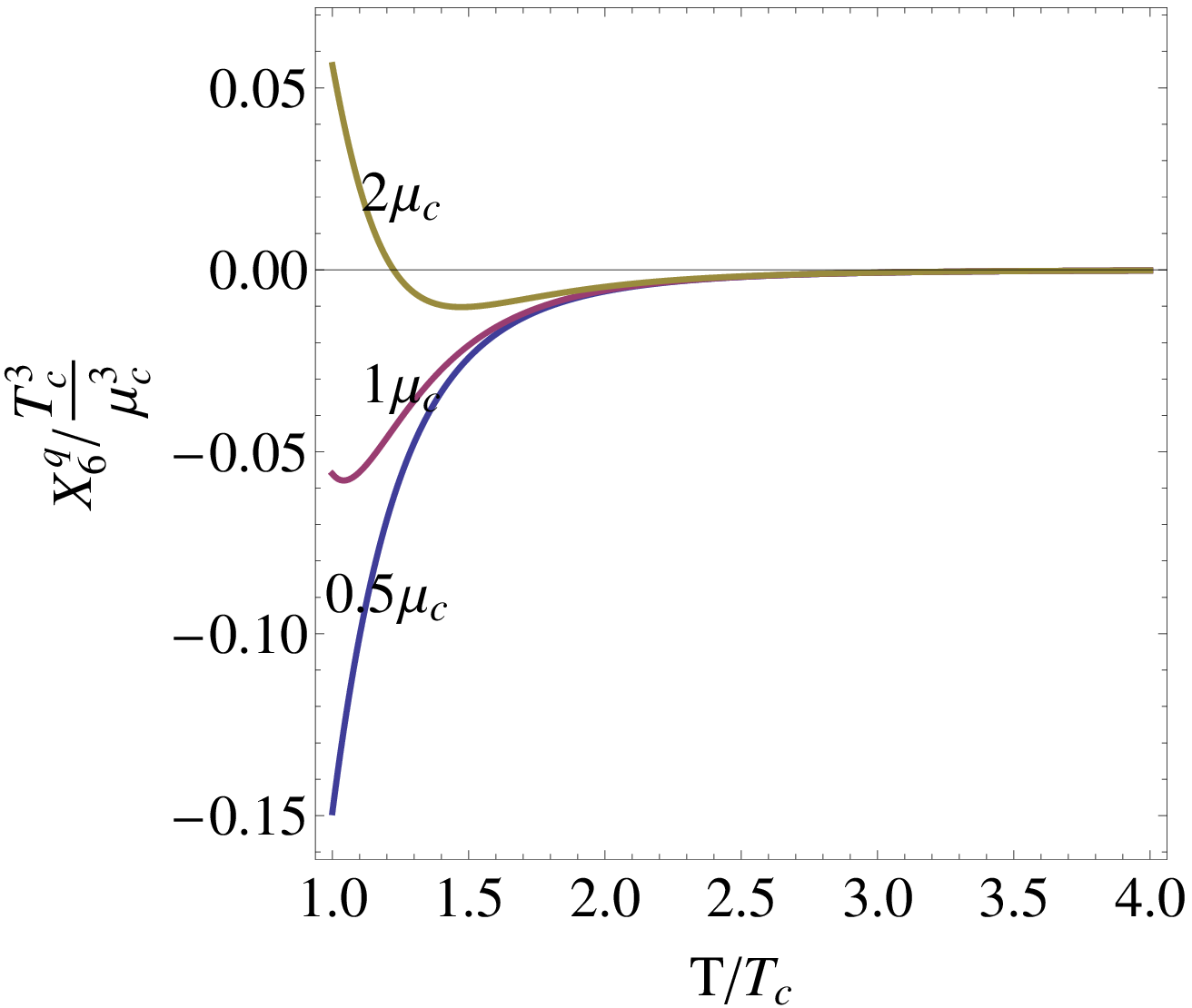} \; 
\caption{ $X^q_{n=1,2,3,4,5,6}$ in unit of $(T_c/\mu_c)^{n-3}$ versus $T/T_c$ at various values of $\mu$ in the QGP phase of the D4/D8 model. The inset at top left conner of $X^q_2$ panel shows the details close to $T_c$.}
\label{fig:d48qgpx}
\end{center}
\end{figure}

\begin{figure}[!h]
\begin{center}
\includegraphics[width=0.45\textwidth]{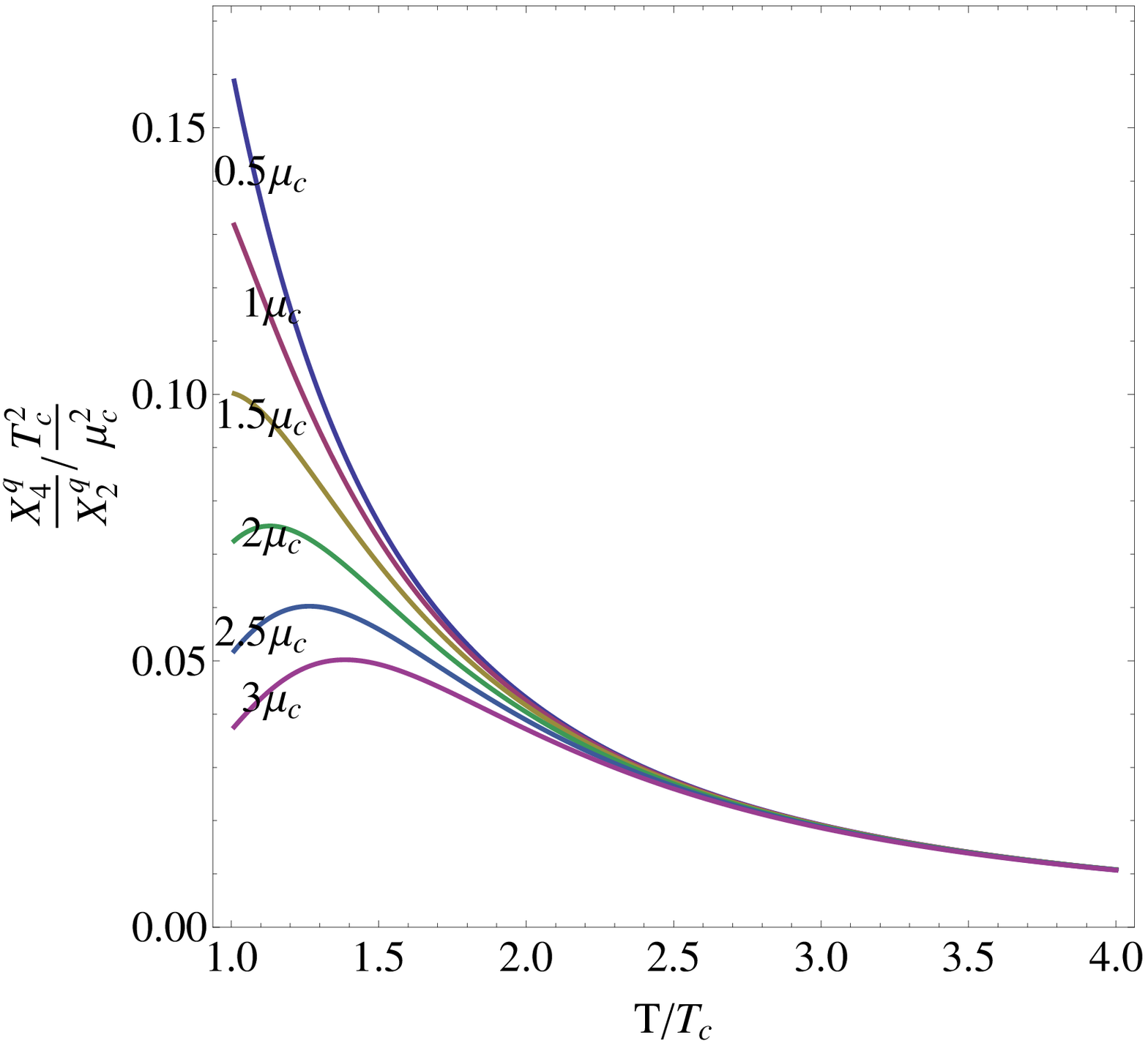}\; 
\includegraphics[width=0.45\textwidth]{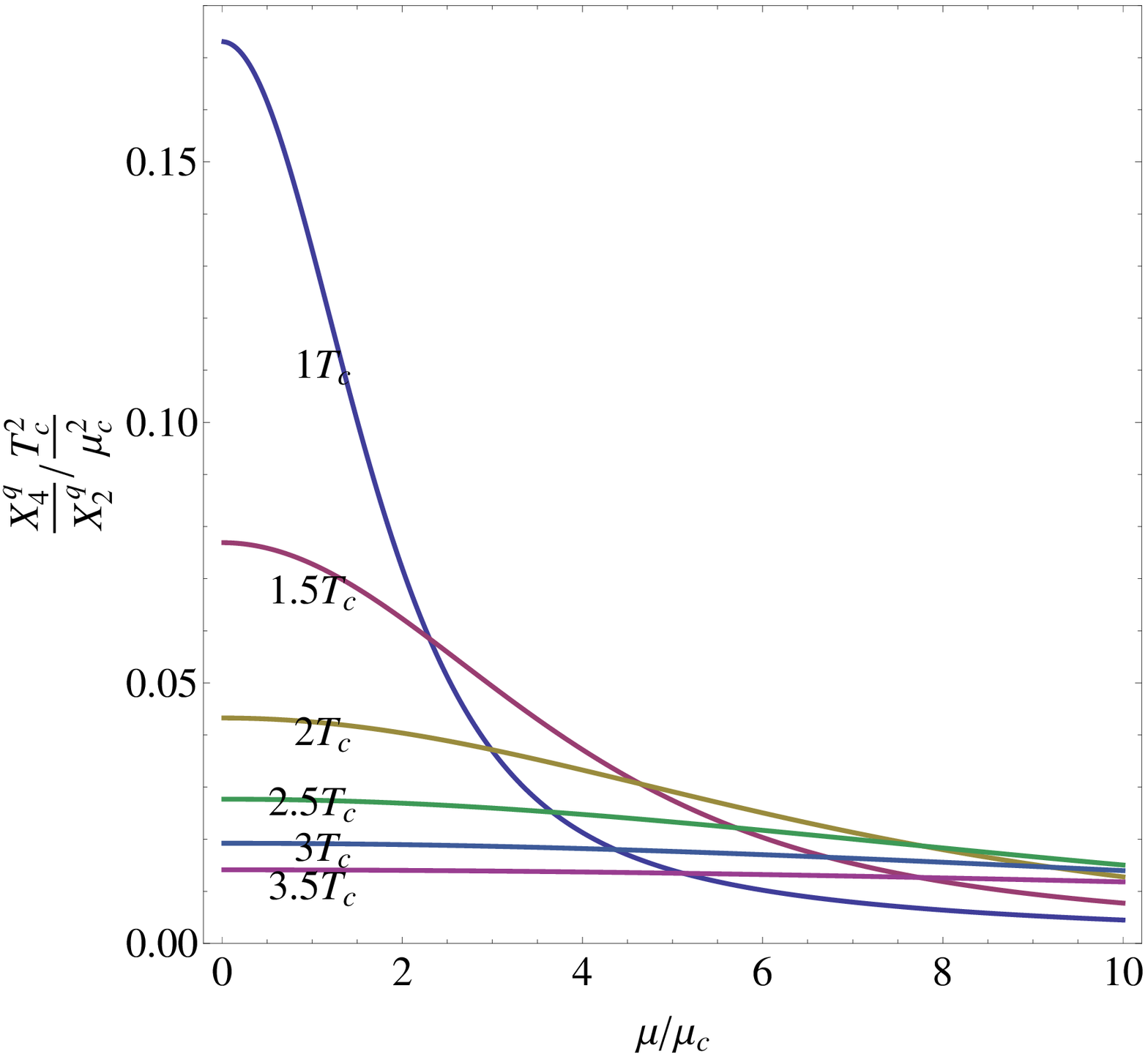}
\caption{$\frac{X^q_4}{X^q_2}$ versus $T$ at various values of $\mu$ (left) and versus $\mu$ at various values of $T$ (right) in the QGP phase of the D4/D8 model.}
\label{fig:sTdm}
\end{center}
\end{figure}

Let us first discuss the results at $\mu=0$ for which only even-order susceptibilities survive. In this case we have simple analytic results:
\begin{eqnarray}
\chi_n^{q} (T) =  \frac{X_n^q(T, \mu=0) }{N_f}= \xi_n \lambda^{3-n}  \left(\frac{T}{T_c}\right)^{3-n}
\end{eqnarray}
with $\xi_n$ being numerical constants, e.g. $\xi_2\approx 0.071$, $\xi_4 \approx 2.2$, and $\xi_6\approx - 158$ for $N_c=N_f=3$. These results were previously obtained in \cite{Kim:2009uu}.

We then examine the patterns of $X^q_n$ at nonzero $\mu$. In Fig.\ref{fig:d48qgpx} we plot $X^q_{n=1,2,3,4,5,6}$ as a function of $T>T_c$ for a variety of values $\mu$. The asymptotic behavior of the first and second order susceptibilities could be understood as follows: analytically one can show that at high T limit one has the leading contribution $P \sim \mu^2 T^3$ and therefore $X^q_1 \sim n_B/T^3 \propto \mu\, T^0$ while $X^q_2 \propto \mu^0\, T$. All the higher order susceptibilities also show interesting patterns and in particular all are suppressed with increasing temperature.  We emphasize that these results are useful and provide direct information on the fluctuation patterns in a strongly interacting matter at regimes with $\mu$ comparable or lager than $T$ where the Taylor expansion via small $\mu/T$ is not useful. 

Let us now discuss the ratio $\frac{X^q_4}{X^q_2}$ (in unit of $T_c^2/\mu_c^2 = (27/(2\lambda))^2 $) which is a quantity of particular interest for the CEP search for a wide range of  $T$ and $\mu$ values in the QGP phase: see Fig.\ref{fig:sTdm}. As one can see from the plots, the general trend is that this ratio decreases when getting to larger $T$ and $\mu$ values and is always positive. If one takes the phase boundary in this model (i.e. the $T=T_c$ line) as an analog to the freeze-out boundary in heavy ion collisions, then walking from small $\mu$ toward large $\mu$  one sees a reducing value for $\frac{X^f_4}{X^f_2}$ which shows a qualitatively similar trend to model computations of QCD and to preliminary STAR measurements.

\subsection{Susceptibilities in the cold dense phase}

In this subsection we study the susceptibilities in the cold dense phase of D4/D8 model. For this phase, the pressure and chemical potential are given by the following equations: 
\begin{eqnarray}
\widetilde P=\int_{1}^{\infty}\Big[1-\frac{1}{\sqrt{1+\frac{\widetilde d^2}{\widetilde u^5}}}\Big]\frac{\widetilde u^4}{\sqrt{\widetilde u^3-1}} \mathrm{d} \widetilde u
\end{eqnarray}
\begin{eqnarray}
\widetilde \mu =1+3 \int_{1}^{\infty}\sqrt{\frac{\widetilde d^2}{\widetilde u^5+\widetilde d^2}}\frac{\widetilde u^{3/2}}{\sqrt{\widetilde u^3-1}} \mathrm{d} \widetilde u
\end{eqnarray}
and both quantities are independent of the temperature. We then compute the susceptibilities with the same definition as in Eq.(\ref{D4D8_Xn}). To do that,  we introduce the following notations: $\widetilde \mu^{(i)}=\frac{\partial^i \widetilde \mu}{\partial \widetilde d^i}$ and $\widetilde P^{(i)}=\frac{\partial^i \widetilde P}{\partial \widetilde d^i}$. By chain rules we have that $\widetilde d = 3\frac{\partial \widetilde P}{\partial \widetilde \mu}=3\frac{\widetilde P^{(1)}}{\widetilde \mu^{(1)}}$, and similarly we can get the following relations:
\begin{eqnarray}
3\widetilde P^{(1)}=\widetilde \mu^{(1)} \widetilde d \; , \;
3\widetilde P^{(2)}=\widetilde \mu^{(2)} \widetilde d+\widetilde \mu^{(1)} \; , \;
3\widetilde P^{(3)}=\widetilde \mu^{(3)} \widetilde d+2\widetilde \mu^{(2)} \; , \;
3\widetilde P^{(4)}=\widetilde \mu^{(4)} \widetilde d+3\widetilde \mu^{(3)}
\end{eqnarray}
With these relations we can then easily derive
%%%%%%%%%%%%%%%%
%%%%%%%%%%%%%%%%
\begin{eqnarray}
\frac{\partial\widetilde P}{\partial \widetilde\mu}&=&\frac{\widetilde d}{3},\\
\frac{\partial^2\widetilde P}{\partial \widetilde\mu^2}&=&\frac{1}{3\widetilde \mu^{(1)}},\\
\frac{\partial^3\widetilde P}{\partial \widetilde\mu^3}&=&-\frac{\widetilde \mu^{(2)}}{3(\widetilde \mu^{(1)})^3},\\
\frac{\partial^4\widetilde P}{\partial \widetilde\mu^4}&=&\frac{3(\widetilde \mu^{(2)})^2-\widetilde \mu^{(1)}\widetilde \mu^{(3)}}{3(\widetilde \mu^{(1)})^5} \; , \;  ...
\end{eqnarray}
where the $\widetilde \mu^{(i)}$ are given by the integrals: 
\begin{eqnarray}
\widetilde \mu^{(i)} = 3 \int_{1}^{\infty}  \frac{\partial^i [(1+ \widetilde u^5 / \widetilde d^2 )^{-1/2}]}{\partial (\widetilde d)^i}      \frac{\widetilde u^{3/2}}{\sqrt{\widetilde u^3-1}} \mathrm{d} \widetilde u
\end{eqnarray}

\begin{figure}[!h]
\begin{center}
\includegraphics[width=0.3\textwidth]{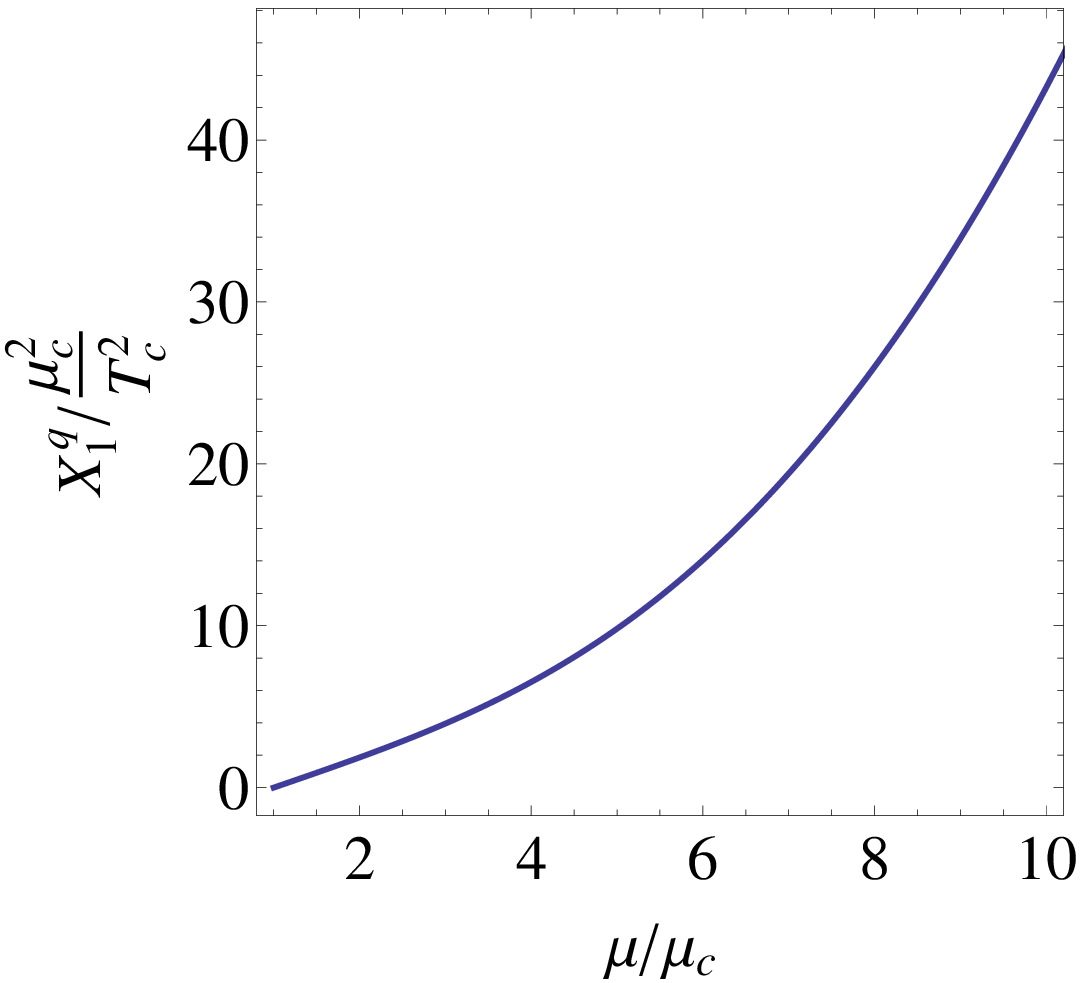} \; 
\includegraphics[width=0.3\textwidth]{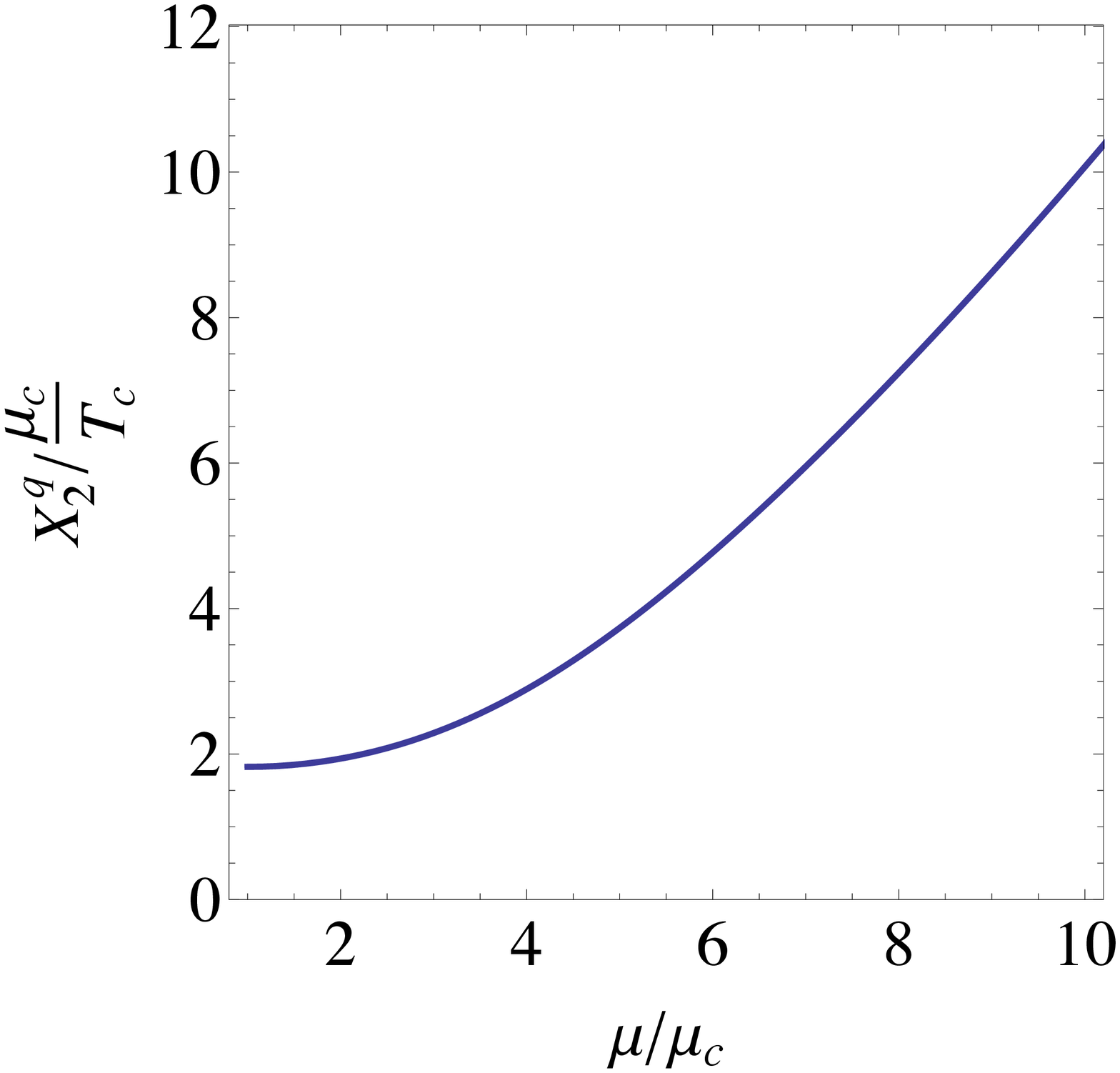} \;
\includegraphics[width=0.29\textwidth]{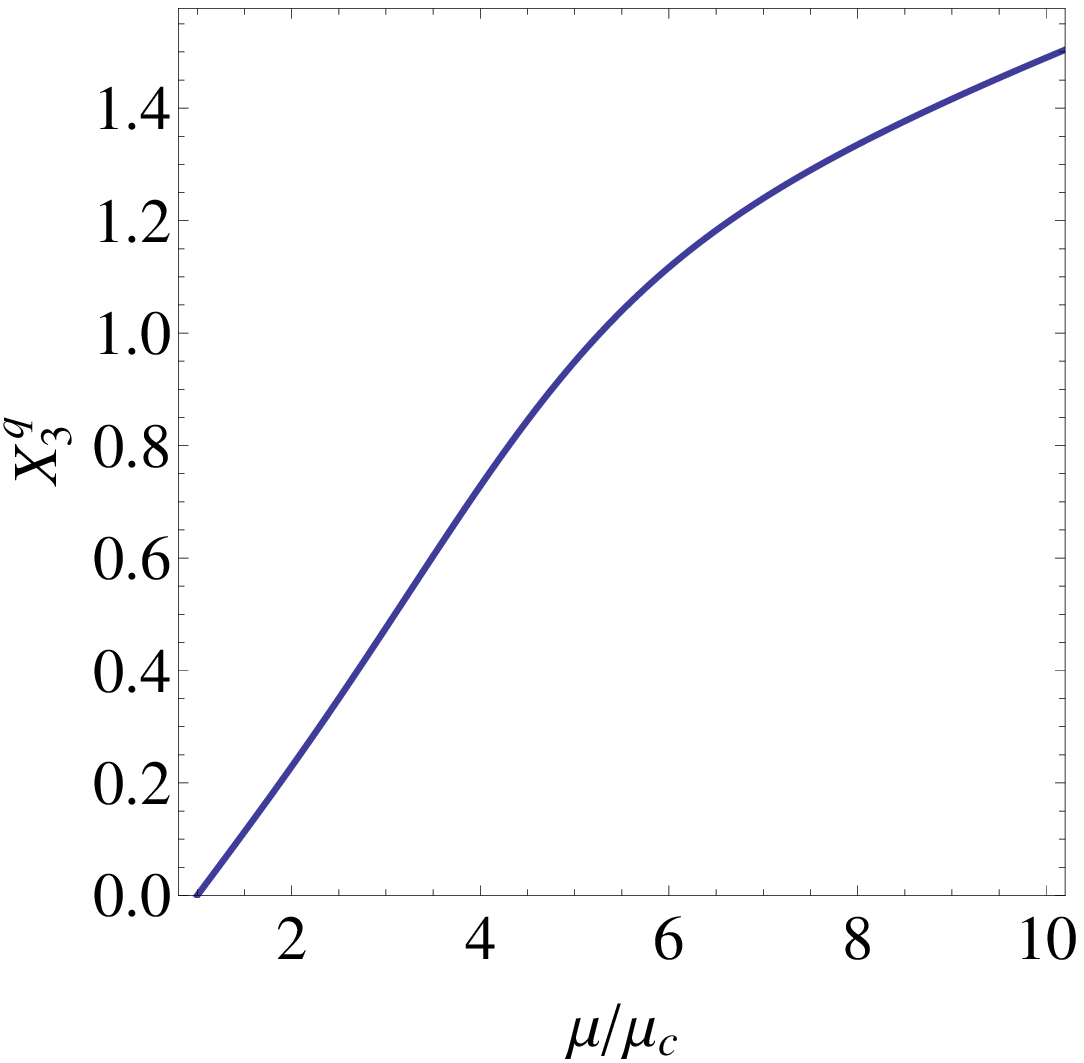} \\
\includegraphics[width=0.30\textwidth]{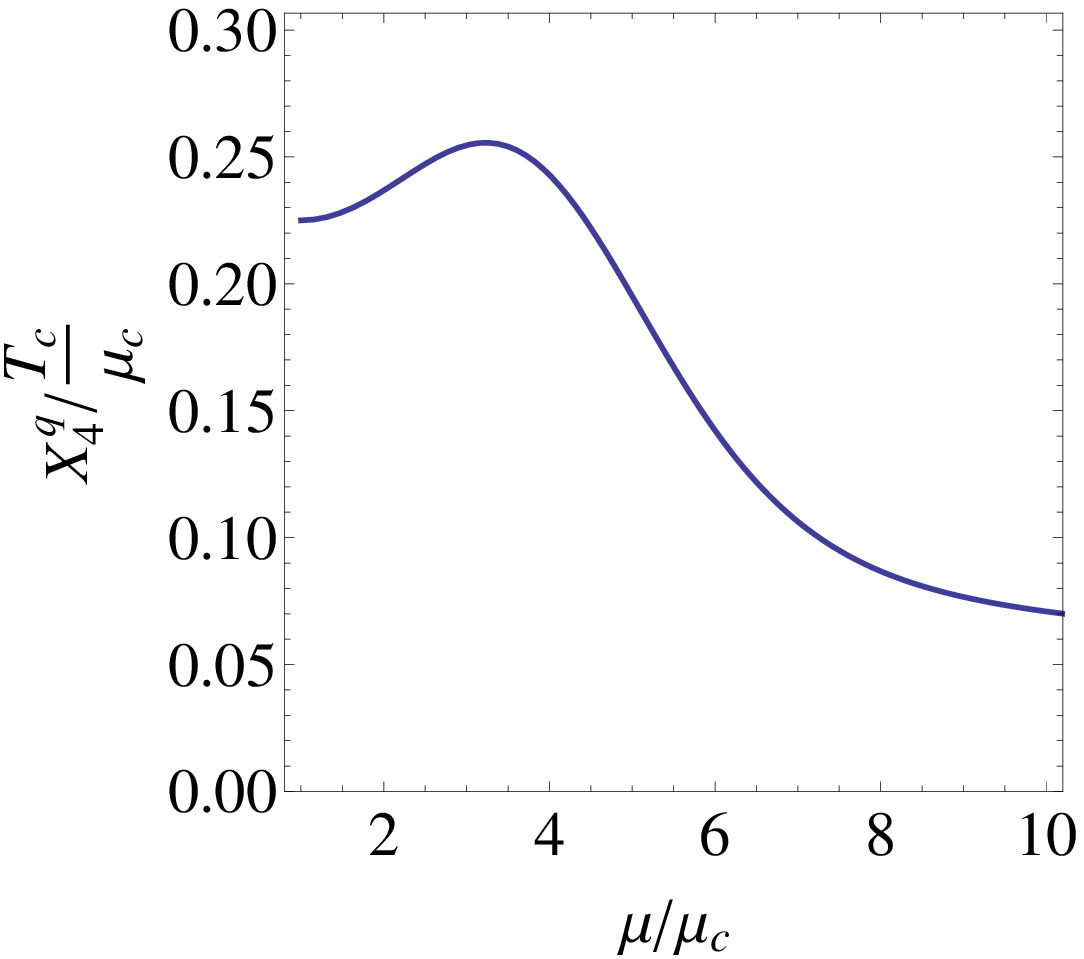} \; 
\includegraphics[width=0.32\textwidth]{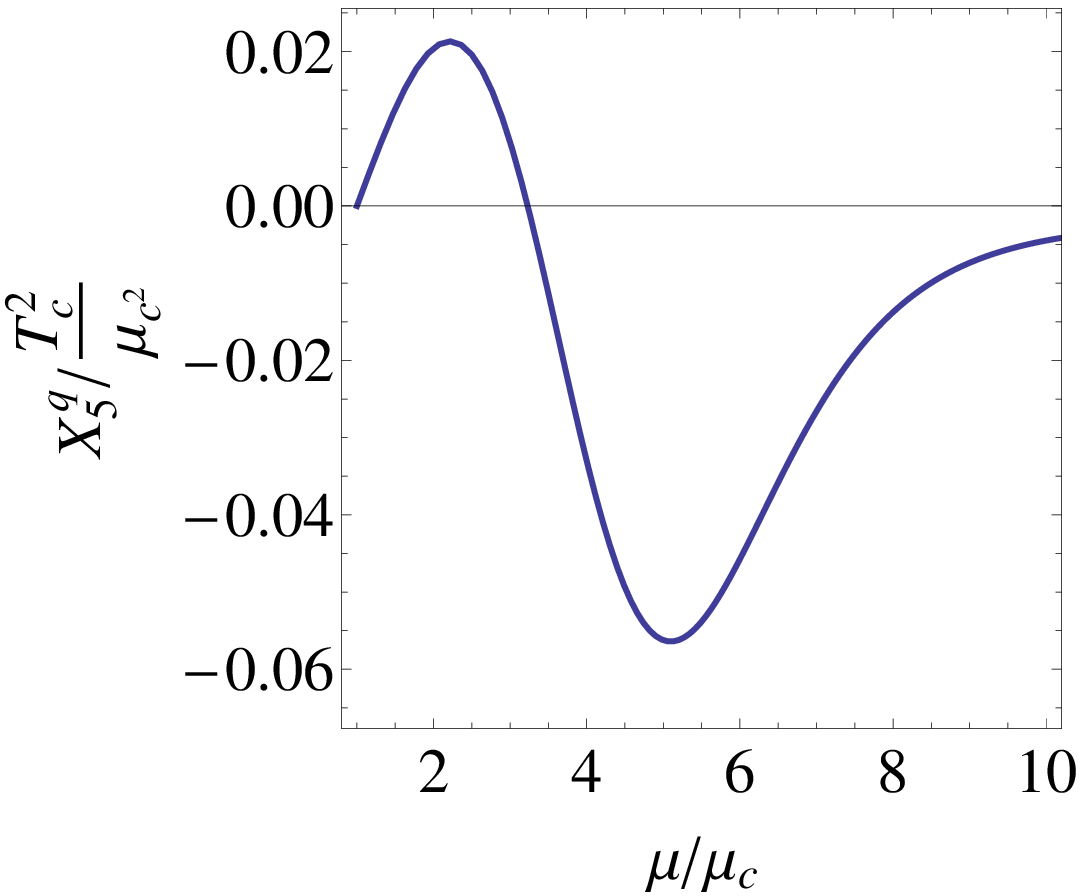} \; 
\includegraphics[width=0.32\textwidth]{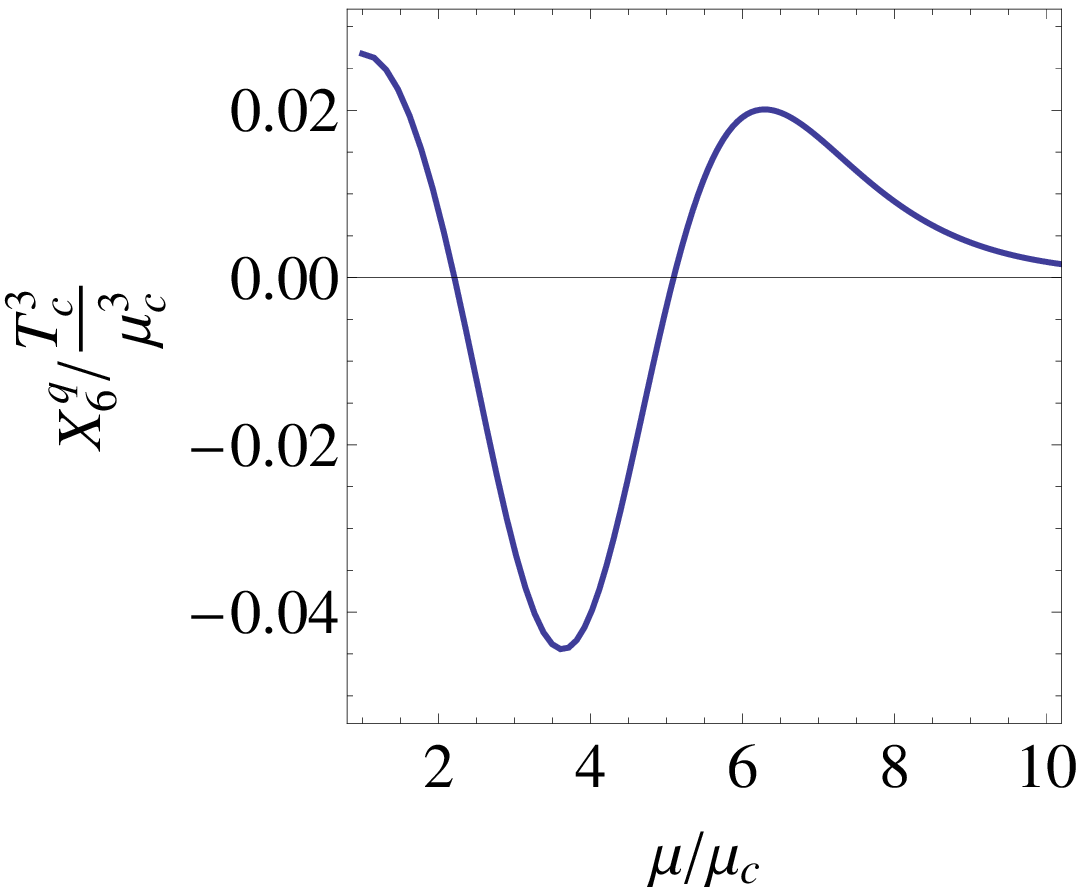} \; 
\caption{ $X^q_{n=1,2,3,4,5,6}$ in unit of $(T_c/\mu_c)^{n-3}$ versus $\mu$ at $T=T_c$ in the cold dense phase of the D4/D8 model.}
\label{fig:d48cdx}
\end{center}
\end{figure}

In Fig.\ref{fig:d48cdx} we show these susceptibilities $X^q_{n=1,2,3,4,5,6}$ as a function of $\mu$. While the first three susceptibilities appear to monotonically grow with chemical potential, the higher order ones get suppressed at large $\mu$. The asymptotic behavior could be qualitatively understood analytically through the following: at large density the density and chemical potential is roughly related as $d \sim \mu^{5/2}$ and thus $X^q_n\sim \mu^{7/2-n}$ so we see the $X^q_{1,2,3}$ to be positive and growing, $X^q_4\sim \mu^{-1/2}$ and $X^q_6\sim \mu^{-5/2}$ to be positive but decreasing, while $X^q_{5}\sim - \mu^{-3/2}$ to be negative and decreasing in magnitude.

For the ratio $\frac{X^q_4}{X^q_2}$ (in unit of $T_c^2/\mu_c^2 = (27/(2\lambda))^2 $), we have the following formulae:
\begin{displaymath}
\frac{X^q_4}{X^q_2} = \frac{3^6}{2^2\lambda^2} \frac{9(\int_{1}^{\infty}\frac{d \widetilde u^{13/2}}{\sqrt{\widetilde u^3-1}(\widetilde u^5+\widetilde d^2)^{5/2}} \mathrm{d} \widetilde u)^2+ \int_{1}^{\infty}\frac{\widetilde u^{13/2}}{\sqrt{\widetilde u^3-1}(\widetilde u^5+\widetilde d^2)^{3/2}} \mathrm{d} \widetilde u \int_{1}^{\infty}\frac{(\widetilde u^5-4\widetilde d^2)\widetilde u^{13/2}}{\sqrt{\widetilde u^3-1}(\widetilde u^5+\widetilde d^2)^{7/2}} \mathrm{d} \widetilde u}{3(\int_{1}^{\infty}\frac{\widetilde u^{13/2}}{\sqrt{\widetilde u^3-1}(\widetilde u^5+\widetilde d^2)^{3/2}} \mathrm{d} \widetilde u)^4}
\end{displaymath}
The results are shown as a function of $\mu$ (noting that there is no $T$-dependence in this cold dense pahse) in Fig.\ref{fig:comp}. For comparison we also show the same ratio in the QGP phase as a function of $\mu$ at $T=T_c$. While both curves show similar magnitude and similar decreasing trend toward large $\mu$, there is a quantitative discontinuity across the phase boundary. For example as labeled in the figure, this ratio at $(T_c,\mu_c)$ is $0.133$ in QGP phase and $0.123$ in cold dense phase though the difference is rather small.  

\begin{figure}[!h]
\begin{center}
\includegraphics[width=0.45\textwidth]{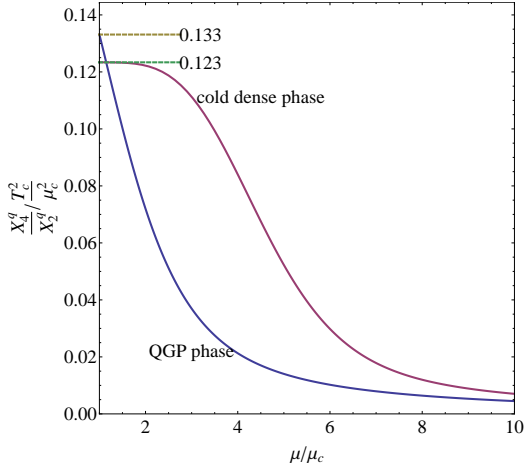}
\caption{ $\frac{X^q_4}{X^q_2}$  versus chemical potential $\mu$ in the cold dense phase of the D4/D8 model  and comparison with the same quantity at $T=T_c$ from the QGP phase of the D4/D8 model.}
\label{fig:comp}
\end{center}
\end{figure}

\section{Susceptibilities from holography: D3/D7 model}

In this section we will evaluate the susceptibilities from another holographic model for QCD based on the D3/D7 branes configuration \cite{Mateos:2006nu,Nakamura:2007nx,Kim:2010zg}. In this model, the background geometry is provided by the $N_c$ D3 branes representing the gauge field dynamics while the fundamental matter is represented by $N_f<<N_c$ (``probe limit'') flavors of D7 branes embedded in the background geometry. These flavor branes are placed at a distance of $M_q$ in the holographic dimension from the D3 branes and the flavor chemical potential $\mu$ is implemented via a $U(1)$ gauge field in the bulk. Different embeddings of the flavor branes correspond to different phases of the model. On the finite $T$ and $\mu$ phase diagram as shown in Fig.\ref{fig:phase} (right panel), there are two distinctive phases: one at low $T$ and $\mu$ corresponding to a ``Minkowski embedding'' with zero density, while the other at high $T$ and $\mu$ corresponding to a ``black hole embedding'' with nonzero density, with the two separated by a phase transition line. In this study we focus on the susceptibilities in the nonzero density phase that has nontrivial dependence on the chemical potential $\mu$. We will closely follow the setup and conventions in \cite{Mateos:2006nu} where one can also find a detailed treatment of the D3/D7 model. For later convenience we also introduce the following definitions and conventions from this model: 
%%%%%%%%%%%%
%%%%%%%%%%%%
\begin{eqnarray}
\widetilde T =\frac{T}{T_c}=\frac{T}{0.764 \times 2 M_q/\sqrt{\lambda}} \; , \; 
\widetilde \mu =\frac{\mu}{M_q} \; , \;  
\widetilde d =\frac{d}{2\pi l_s^2 u_0^3 N_f T_{D7}} \; , \; 
\widetilde P =\frac{P}{4\pi^3 l_s^2 u_0^3 N_f T_{D7}M_q} \; .
\end{eqnarray}
with parameters $l_s,g_s$ being the string length and coupling, $\lambda=g^2N_c=2\pi g_s N_c$ the 't Hooft coupling, $u_0$ the horizon parameter and $T_{D7}$ the tension of the D7 branes (see more details in \cite{Mateos:2006nu}). Here we study the finite density phase above the transition line corresponding to the black hole embedding.

The susceptibilities we would like to calculate are defined as
%%%%%%%%%%%
%%%%%%%%%%%
\begin{eqnarray} \label{D3D7_chi}
X^q_n(T, \mu)=\frac{\partial^n(P/T^4)}{\partial (\mu/T)^n}=\left (2^{n-7/2} 0.764^{n-1} \right )  N_c N_f \lambda^{1-\frac{n}{2}} \widetilde T^{n-1}\frac{\partial^n\widetilde P}{\partial \widetilde\mu^n}
\end{eqnarray}
with the relation between the   chemical potential $\mu$ and density $d$ determined by 
\begin{eqnarray} \label{D3D7_mu}
\widetilde \mu=2 \sqrt{2} \times 0.764\widetilde T \widetilde d \int^{\infty}_{1}\mathrm{d}\rho\frac{f\sqrt{1-\chi^2+\rho^2\dot{\chi}^2}}{\sqrt{\widetilde f(1-\chi^2)[\rho^6\widetilde f^3(1-\chi^2)^3+8\widetilde d^2]}}
\end{eqnarray}
In the above the factors $f=1-\rho^{-4}$ and $\widetilde f=1+\rho^{-4}$ are from metric (with $\rho$ the radial coordinate). The function $\chi(\rho)$ describes the embedding of the D7 branes in the geometric background (with $\dot{\chi}=d\chi/d\rho$) and is solved from the equation of motion below\cite{Mateos:2006nu}:
\begin{eqnarray} \label{D3D7_eom}
&&\partial_{\rho} \Bigg[ \frac{\rho^5 f \widetilde f (1-\chi^2) \dot{\chi}}{\sqrt{1-\chi^2+\rho^2\dot{\chi}^2}} \sqrt{1+\frac{8 \widetilde d^2}{\rho^6 \widetilde f^3(1-\chi^2)^3}} \Bigg] \nonumber \\
&& \quad =  - \frac{\rho^3 f \widetilde f \chi}{\sqrt{1-\chi^2+\rho^2\dot{\chi}^2}} \sqrt{1+\frac{8 \widetilde d^2}{\rho^6 \widetilde f^3(1-\chi^2)^3}} \Bigg[ 3 (1-\chi^2) + 2 \rho^2 \dot{\chi}^2 - 24 \widetilde d^2 \frac{1-\chi^2+\rho^2 \dot{\chi}^2}{\rho^6 \widetilde f^3(1-\chi^2)^3 +8 \widetilde d^2} \Bigg]
\end{eqnarray} 
One can determine the phase boundary $(T,\mu)_c$ by solving the Eqs.(\ref{D3D7_mu},\ref{D3D7_eom}) in the limit of vanishing  density $d \to 0$ as shown in Fig.\ref{fig:phase} (right panel). Here we define $T_c$ and $\mu_c$ as the crosspoint of the phase boundary line and $T,\mu$ axes, respectively. One can get $\mu_c=M_q$ and $T_c=0.764\times \frac{2M_q}{\sqrt{\lambda}}$ as in \cite{Mateos:2006nu}.

Our strategy to compute $X^q_n$ in Eq.(\ref{D3D7_chi}) is to start with density $d$ as $\partial^n P /\partial \mu^n \sim \partial^{n-1} d / \partial \mu^{n-1}$. To do that let us first introduce $\widetilde \mu^{(i)}=\frac{\partial^i \widetilde \mu}{\partial \widetilde d^i}$ and the integrals $I_i$ defined as
\begin{eqnarray}
I_i=\int^{\infty}_{1}\mathrm{d}\rho\frac{f\sqrt{1-\chi^2+\rho^2\dot{\chi}^2}}{\sqrt{\widetilde f(1-\chi^2)}}[\rho^6\widetilde f^3(1-\chi^2)^3+8\widetilde d^2]^{-\frac{2i-1}{2}},\big( i=1,2,...,6 \big)
\end{eqnarray} 
With these integrals, the Eq.(\ref{D3D7_mu}) can be simply written as $\widetilde \mu=2\sqrt{2}\times ( 0.764 \widetilde T)\,  \widetilde d\, I_1$.  One can further obtain the following results by differentiations: 
\begin{eqnarray}
&& \widetilde \mu^{(1)}=\sqrt{2}\times 0.764 \widetilde T (2 I_1 -16 \widetilde d^2 I_2) \; , \; 
\widetilde \mu^{(2)}=\sqrt{2}\times 0.764 \widetilde T (-48\widetilde d I_2 +384 \widetilde d^3 I_3) \; , \nonumber \\
&& \widetilde \mu^{(3)}=\sqrt{2}\times 0.764 \widetilde T (-48 I_2 +2~304 \widetilde d^2 I_3 -15~360 \widetilde d^4 I_4) \; ,\nonumber \\
&& \widetilde \mu^{(4)}=\sqrt{2}\times 0.764 \widetilde T (5~760\widetilde d I_3-153~600\widetilde d^3 I_4 +860~160\widetilde d^5 I_5) \; , \; ...
\end{eqnarray}
%%%%%%%%%%%
%%%%%%%%%%%
Finally we can get the following formulae for the susceptibilities :
\begin{eqnarray}
\frac{\partial\widetilde P}{\partial \widetilde\mu}&=&{\widetilde d},\\
\frac{\partial^2\widetilde P}{\partial \widetilde\mu^2}&=&\frac{1}{\widetilde \mu^{(1)}},\\
\frac{\partial^3\widetilde P}{\partial \widetilde\mu^3}&=&-\frac{\widetilde \mu^{(2)}}{(\widetilde \mu^{(1)})^3},\\
\frac{\partial^4\widetilde P}{\partial \widetilde\mu^4}&=&\frac{3(\widetilde \mu^{(2)})^2-\widetilde \mu^{(1)}\widetilde \mu^{(3)}}{(\widetilde \mu^{(1)})^5} \; , \;  ...
\end{eqnarray}

These susceptibilities can be analytically obtained in the very high temperature limit $T >> M_q$ and vanishing density $d\to 0$, for which one simply has $\chi \to 0$. In this limit we have 
\begin{eqnarray}
I_1 \to \frac{1}{4} \; , \; I_2 \to \frac{1}{128} \; , \; I_3 \to \frac{1}{1792} \; , \; I_4 \to \frac{1}{20480} \; , \; ...
\end{eqnarray}
and can get the following results for susceptibilities: 
%%%%%%%%%
%%%%%%%%%
\begin{eqnarray}
\chi^q_2(T) \to \frac{3}{2}   \; , \;  \chi^q_4(T) \to \frac{9}{\lambda}  \; , \; \chi^q_6(T) \to  - \frac{540}{7}  \frac{1}{\lambda^2} \; , \; ...
\end{eqnarray}

%%%%%%%%%%%%%
%%%%%%%%%%%%%
\begin{figure}[!h]
\begin{center}
\includegraphics[width=0.31\textwidth]{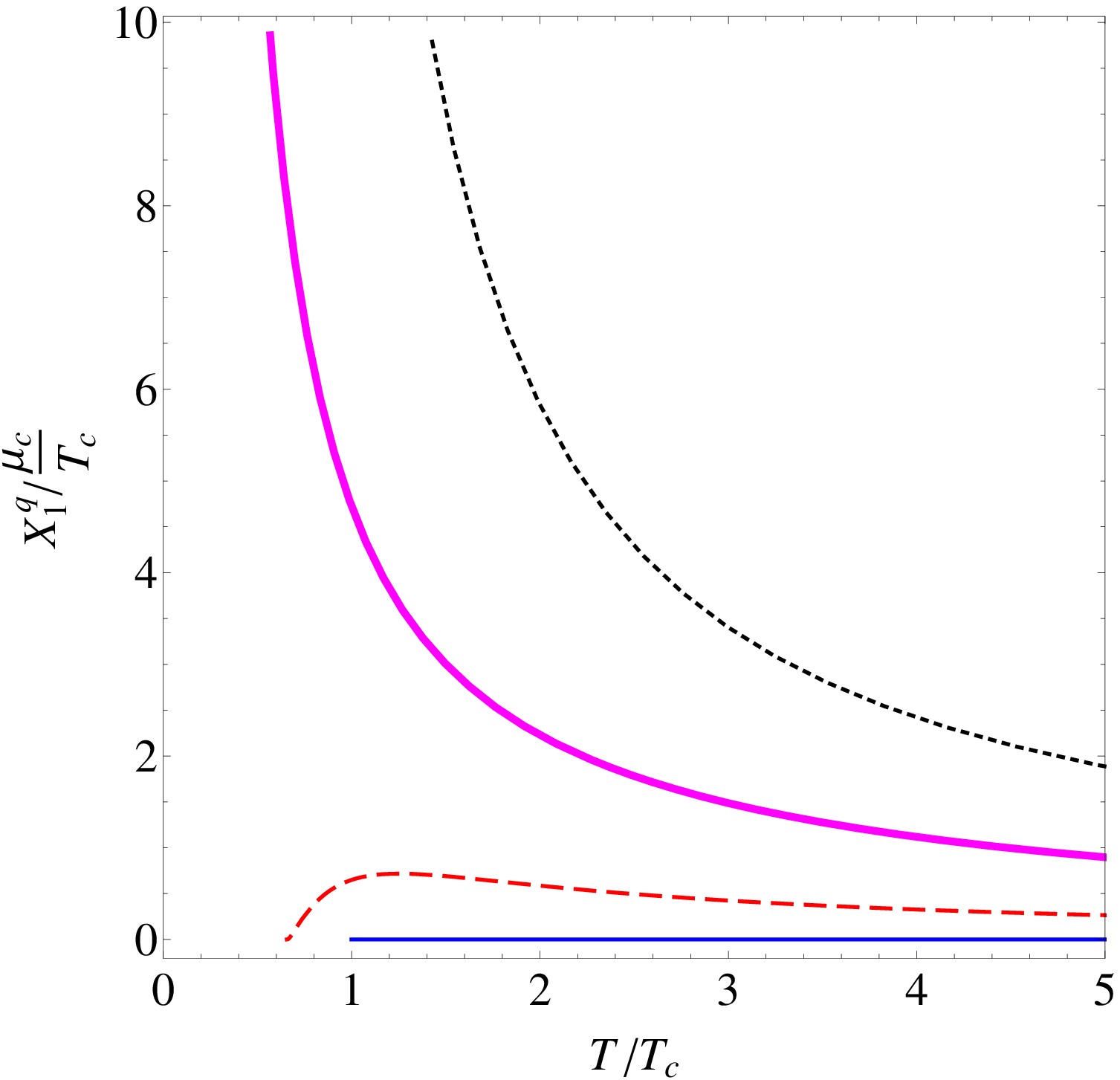} \; 
\includegraphics[width=0.29\textwidth]{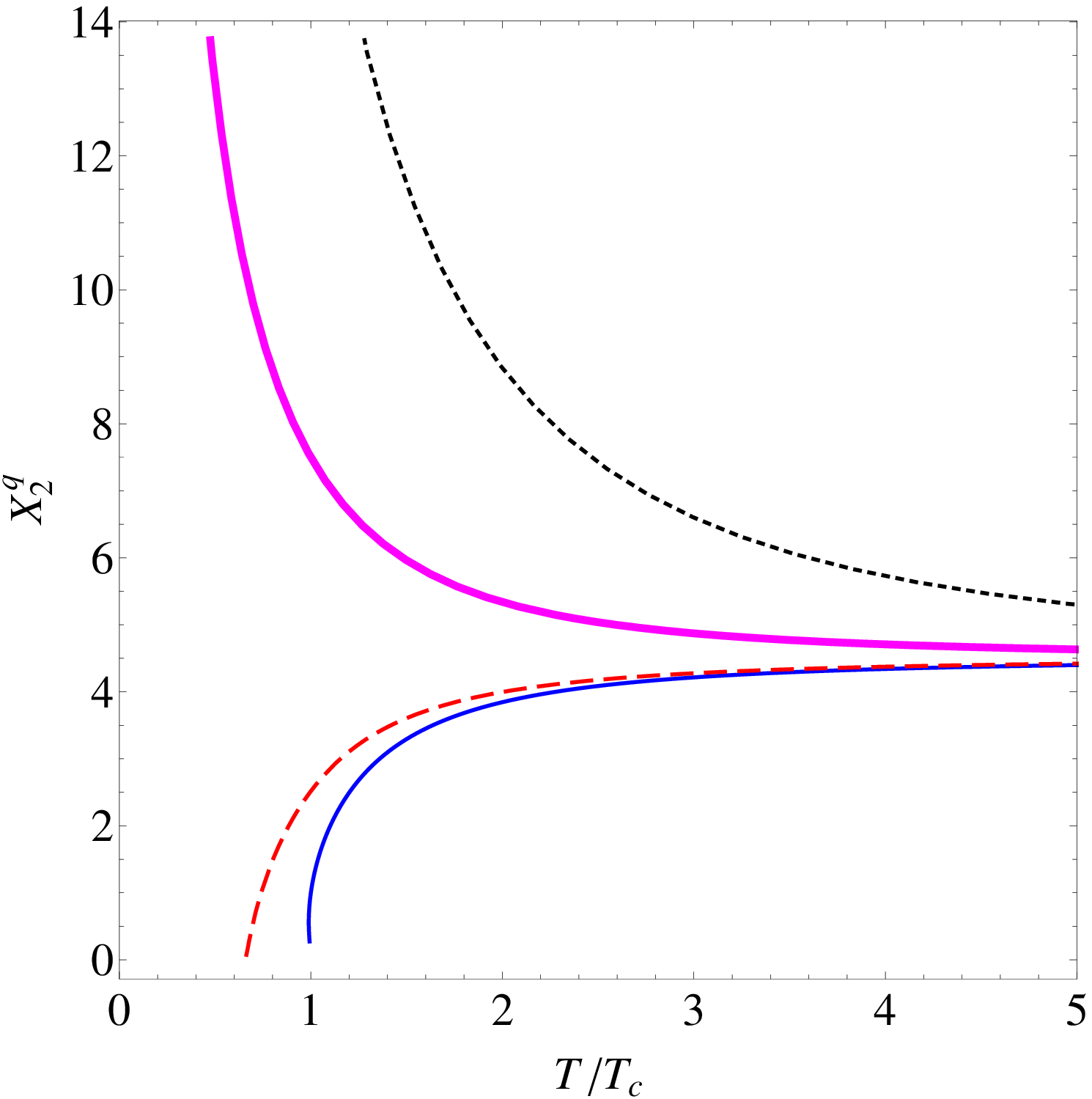} \; 
\includegraphics[width=0.31\textwidth]{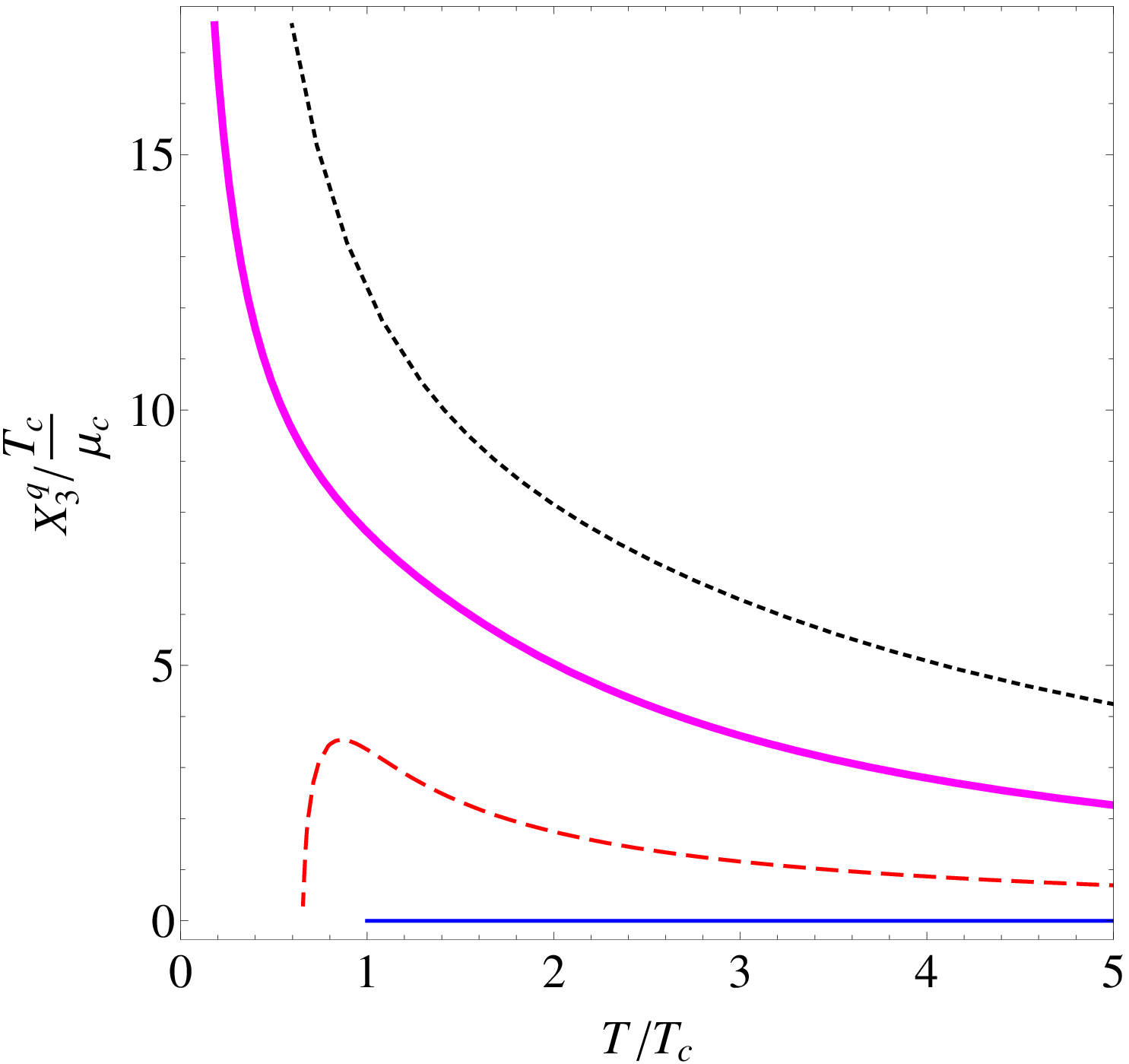} \\
\includegraphics[width=0.30\textwidth]{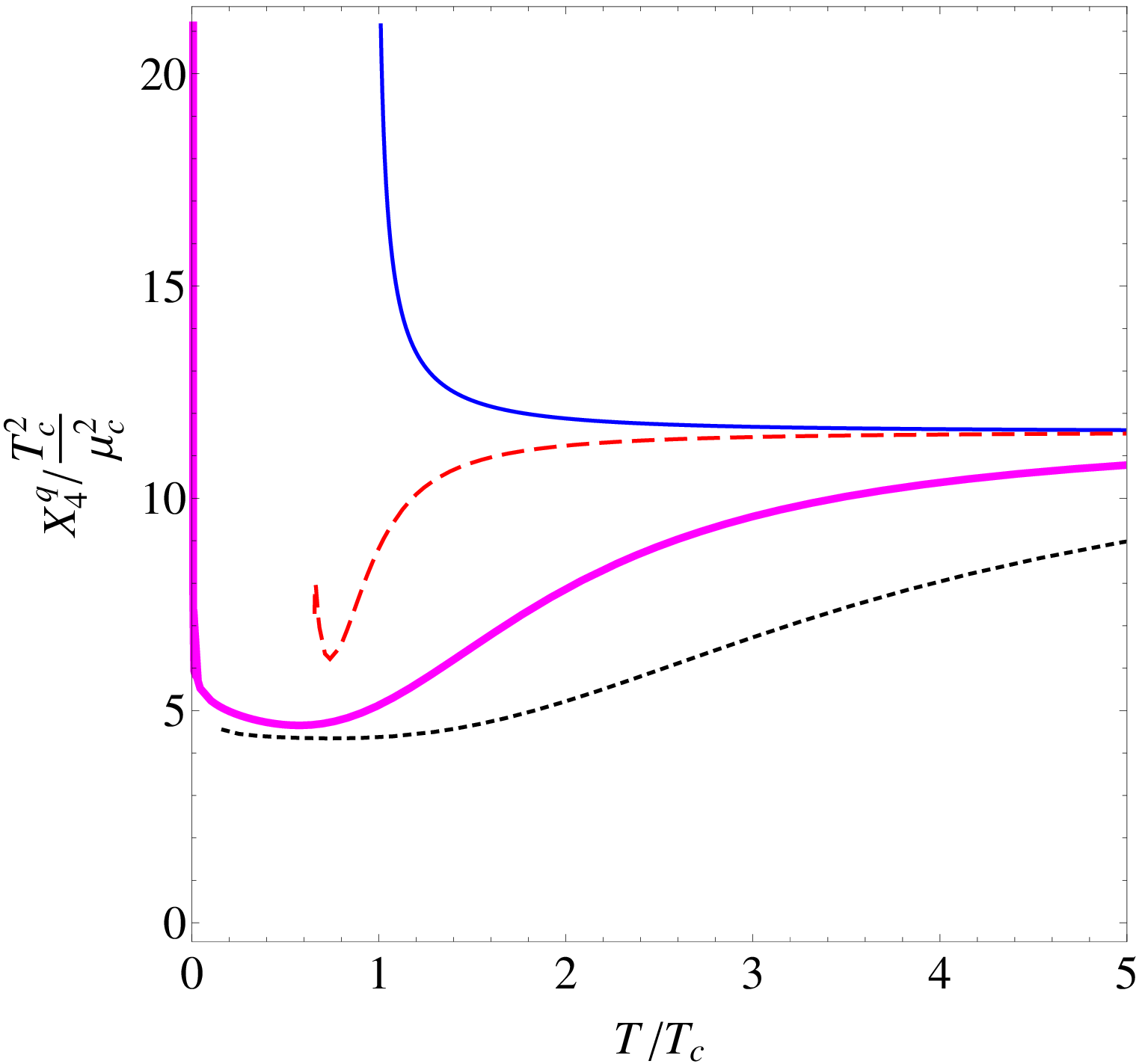} \; 
\includegraphics[width=0.31\textwidth]{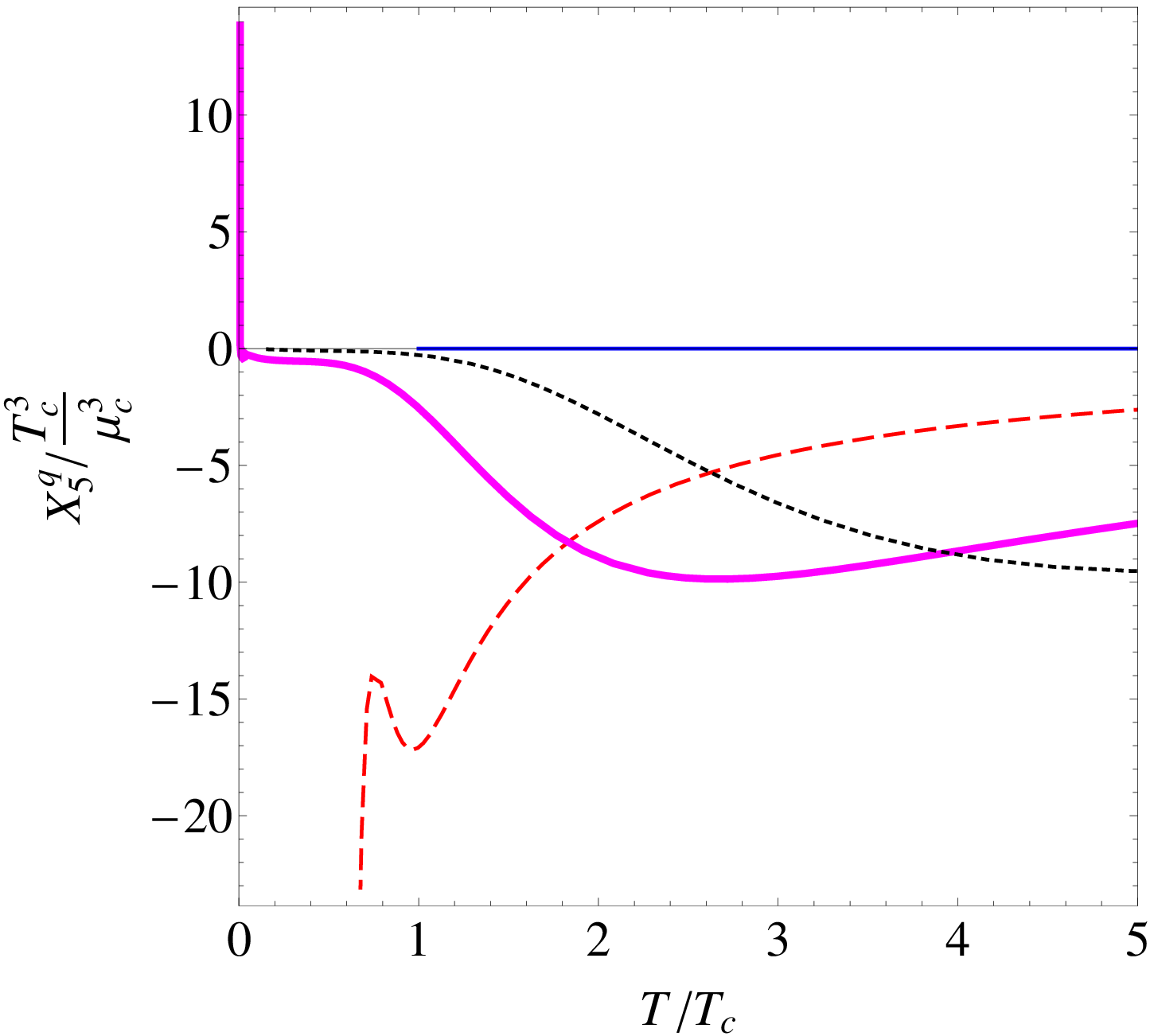} \; 
\includegraphics[width=0.32\textwidth]{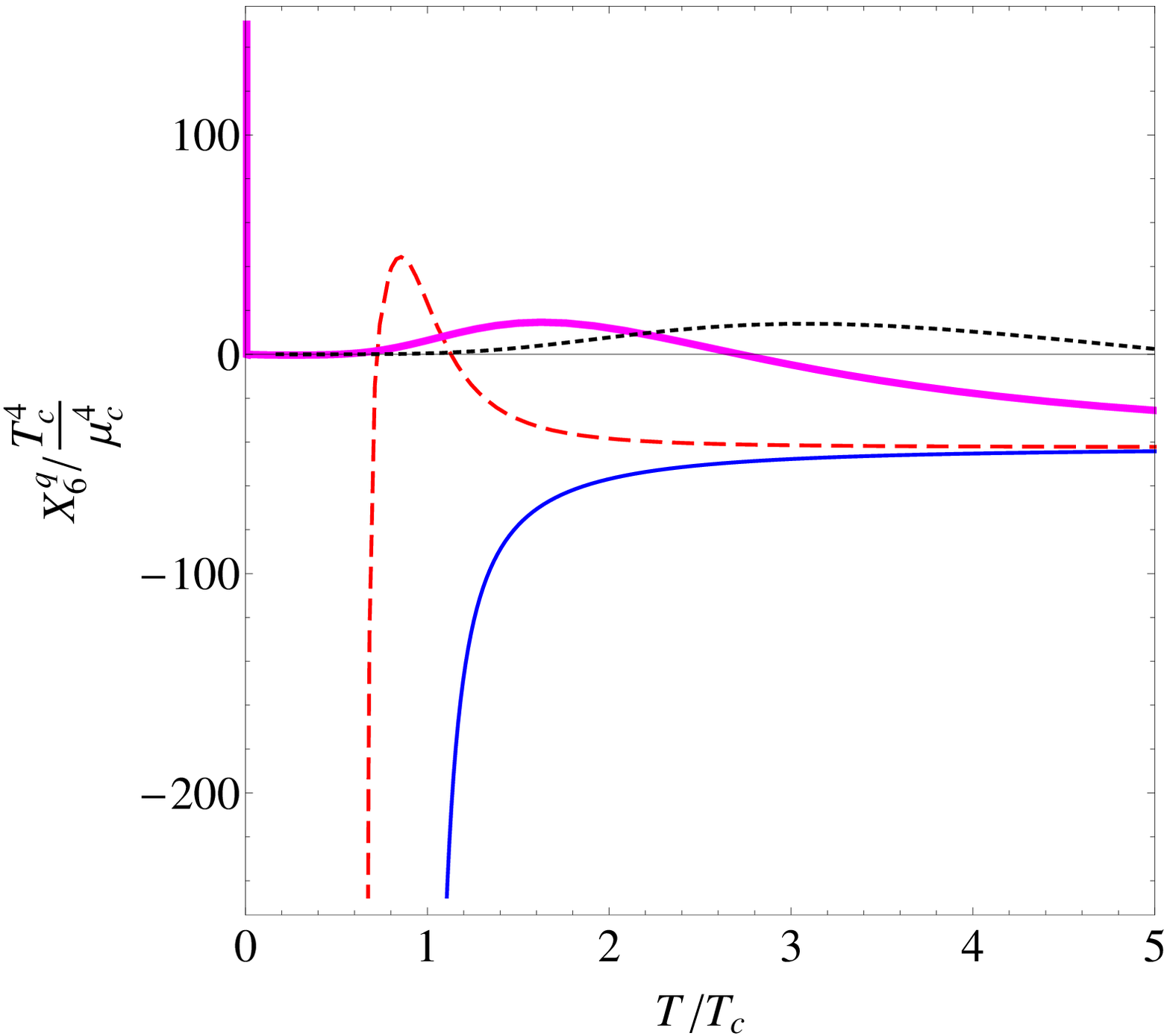} \; 
\caption{ $X^q_{n=1,2,3,4,5,6}$ in unit of $(T_c/\mu_c)^{n-2}$ versus $T$ at  $\mu=0$ (thin solid blue), $\mu=0.3\mu_c$ (dashed red), $\mu=\mu_c$ (thick solid magenta), and $\mu=2\mu_c$ (dotted black) from the D3/D7 model, respectively. }
\label{fig:d37xvsT}
\end{center}
\end{figure}

\begin{figure}[!h]
\begin{center}
\includegraphics[width=0.45\textwidth]{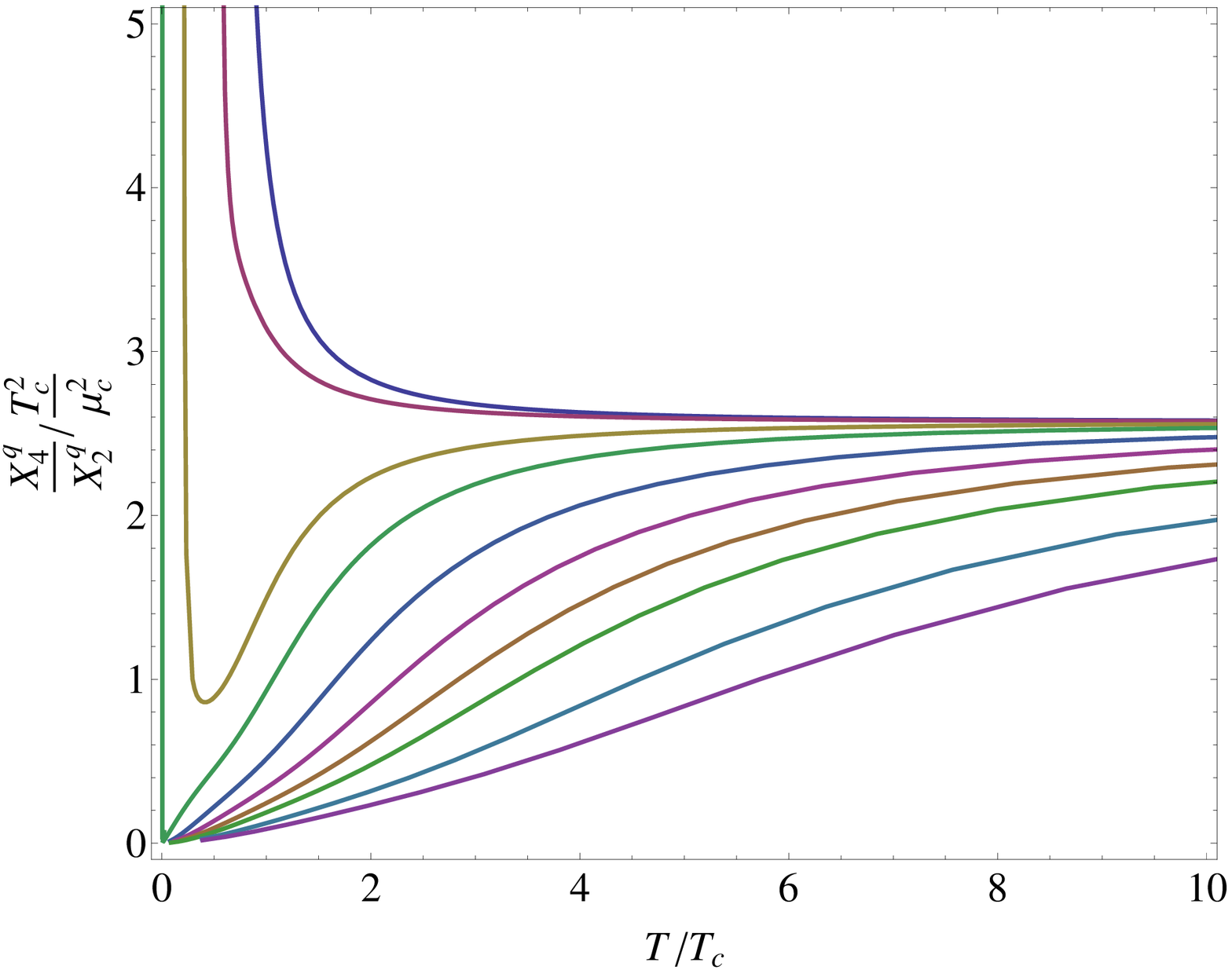}\;
\includegraphics[width=0.45\textwidth]{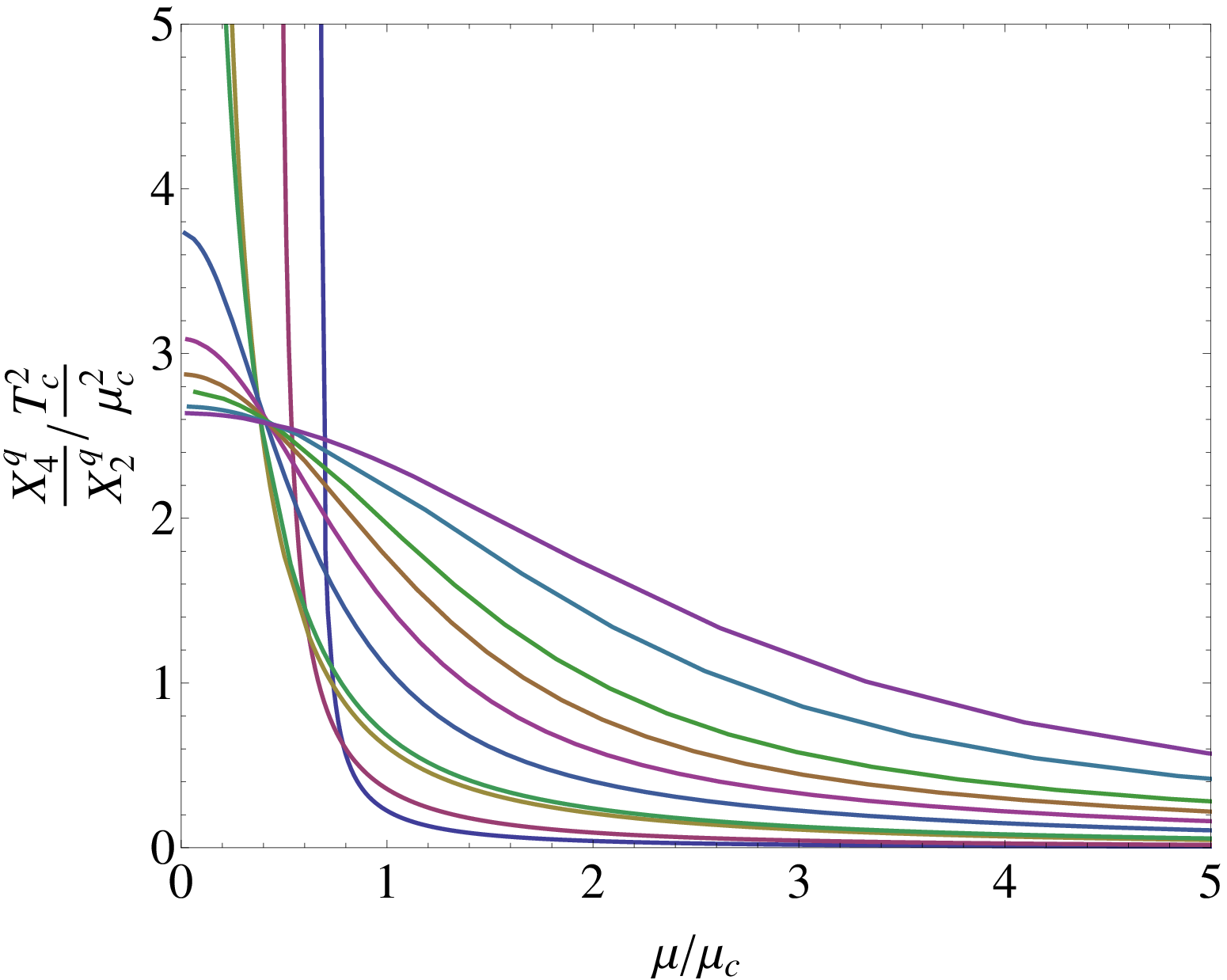}
\caption{
$\frac{X^q_4}{X^q_2}$ versus $T$ at various values of $\mu$ (left, from bottom up with $\widetilde \mu$=5, 4, 3, 2.5, 2, 1.5, 1, 0.7, 0.3, 0.1 ) and versus $\mu$ at various values of $T$ (right, from bottom up with $\widetilde T$=0.3, 0.5, 0.9, 1, 1.5, 2, 2.5, 3, 4, 5) in the D3/D7 model. }
\label{fig:37s}
\end{center}
\end{figure}

Generally for any given temperature and density one can solve the equation of motion for $\chi(\rho)$ numerically and then compute the susceptibilities as in the above formulae. In Fig.\ref{fig:d37xvsT} we plot $X^q_{n=1,2,3,4,5,6}$ as a function of $T$ for a variety of values $\mu$. A first distinction from the D4/D8 model is that the susceptibilities diverge when approaching the phase boundary. A second interesting pattern is that all the susceptibilities appear to approach certain constant values toward large temperature in contrast to the results from D4/D8 model that show strong T-dependence as $T\to \infty$. We will discuss this distinction more in the next Section. Finally we notice that the odd susceptibilities for $\mu=0$ vanish as required by symmetry. 

Finally we discuss the ratio $\frac{X^q_4}{X^q_2}$ (in unit of $T_c^2/\mu_c^2 = 0.764^2 \cdot 4/\lambda $) which is shown for a wide range of  $T$ and $\mu$ values in the D3/D7 phase: see Fig.\ref{fig:37s}. A most distinctive feature here, as compared with the D4/D8 results, is that the ratio diverges when approaching the phase transition line from either $T$ or $\mu$ directions. Also different from the D4/D8 case is that (for non-divergent regime $T>T_c$ and/or $\mu>\mu_c$) the ratio here increases with temperature, though it decreases with $\mu$ in both models.

\section{Discussions in light of current lattice QCD results}

Lattice QCD simulations provide the ultimate answers for the precise values of these susceptibilities in real QCD matter (albeit limited to the very small density region due to the sign problem). To understand the implications of these results, however, could be tricky particularly in the strongly interacting regime. Current lattice QCD results (mostly for the second order susceptibilities) from  both Wuppertal-Budapest group \cite{Borsanyi:2011sw} and BNL-Bielefeld group \cite{Bazavov:2012jq} are at almost physical masses, with high precision, and in agreement with each other. In light of these lattice input,  we discuss in this section what we can learn from various models about susceptibilities and whether these models provide viable and consistent descriptions for one or more aspects of the lattice QCD data. 

\subsection{What can we learn from holographic models?}

With  the susceptibilities being computed from both D4/D8 and D3/D7 models, it is tempting to compare the two and find common as well as different features. Let us focus on the susceptibilities $\chi^q_{n=2,4,6,...}(T)$ at zero density.  

One very interesting {\em common feature} of the two models is that all higher order susceptibilities $n>2$ are suppressed by inverse powers of strong coupling $\lambda$. Namely, we have 
\begin{eqnarray}
\chi^q_n \propto  \lambda^{3-n} \quad (D4/D8) \; \\
\chi^q_n \propto   \lambda^{1-\frac{n}{2}} \quad (D3/D7) \; 
\end{eqnarray}
Clearly in the infinity coupling limit $\lambda\to \infty$ only the second order $\chi^q_2$ survives in both models, which seems to suggest that {\em the fluctuations of conserved charges become Guassian at very strong coupling}. This feature might be universal, and if so would be the most interesting lesson we learn from holographic models. 

We can compare the temperature dependence of $\chi^q_n$ in the two models, as shown in Fig.\ref{fig:comp} for $n=2,4,6$. We also show the lattice results for $\chi_2^u$ (light quark number susceptibility) \cite{Borsanyi:2011sw,Bazavov:2012jq}. These plots are obtained with coupling parameter $\lambda=9$ for both models and one may note that such a choice of $\lambda$ may not be the optimal one for other physics considerations in these models and also the optimal choice of $\lambda$ may not be necessarily the same for the two models. In fact the higher order susceptibilities are rather sensitive to the choice of $\lambda$. While it is hard to consider these results as quantitative explanation of the actual QGP susceptibilities, the qualitative trends in such holographic models appear rather interesting and informative. 

Finally we would like to compare the high temperature limit of the $\chi^q_n$ for both models: 
\begin{eqnarray}
\chi^q_n(T\to \infty) \propto   \left(\frac{T}{T_c} \right)^{3-n}  \quad (D4/D8) \; \\
\chi^q_n(T\to \infty) \propto   \left(\frac{T}{T_c} \right)^{0}  \quad (D3/D7) \; 
\end{eqnarray} 
There are interesting differences between the two models: the D3/D7 model results show constant asymptotic values at high temperature for all orders (as also evident from the plots in Fig.\ref{fig:comp}), which may be understandable as there is essentially only one scale for the model in the  limit $T>>M_q$; the D4/D8 model results however show strong temperature dependence which may be due to the two scales in the model even at high $T$ limit (with more and more high KK models involved), and furthermore at high $T$ all higher order $n>2$ susceptibilities are suppressed and the fluctuation becomes Guassian. 

\begin{figure}[!h]
\begin{center}
\includegraphics[width=0.3\textwidth]{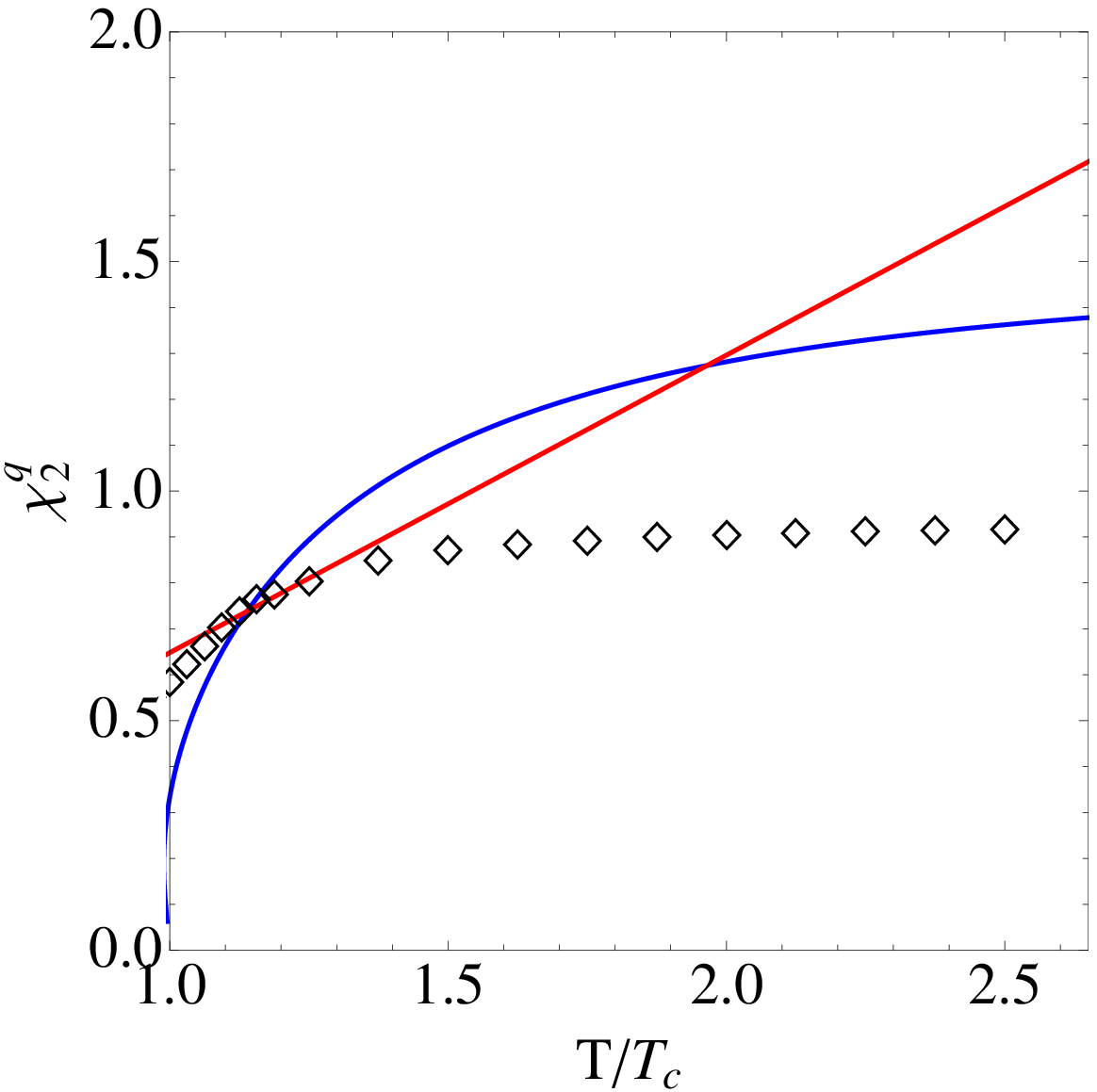} \;
\includegraphics[width=0.3\textwidth]{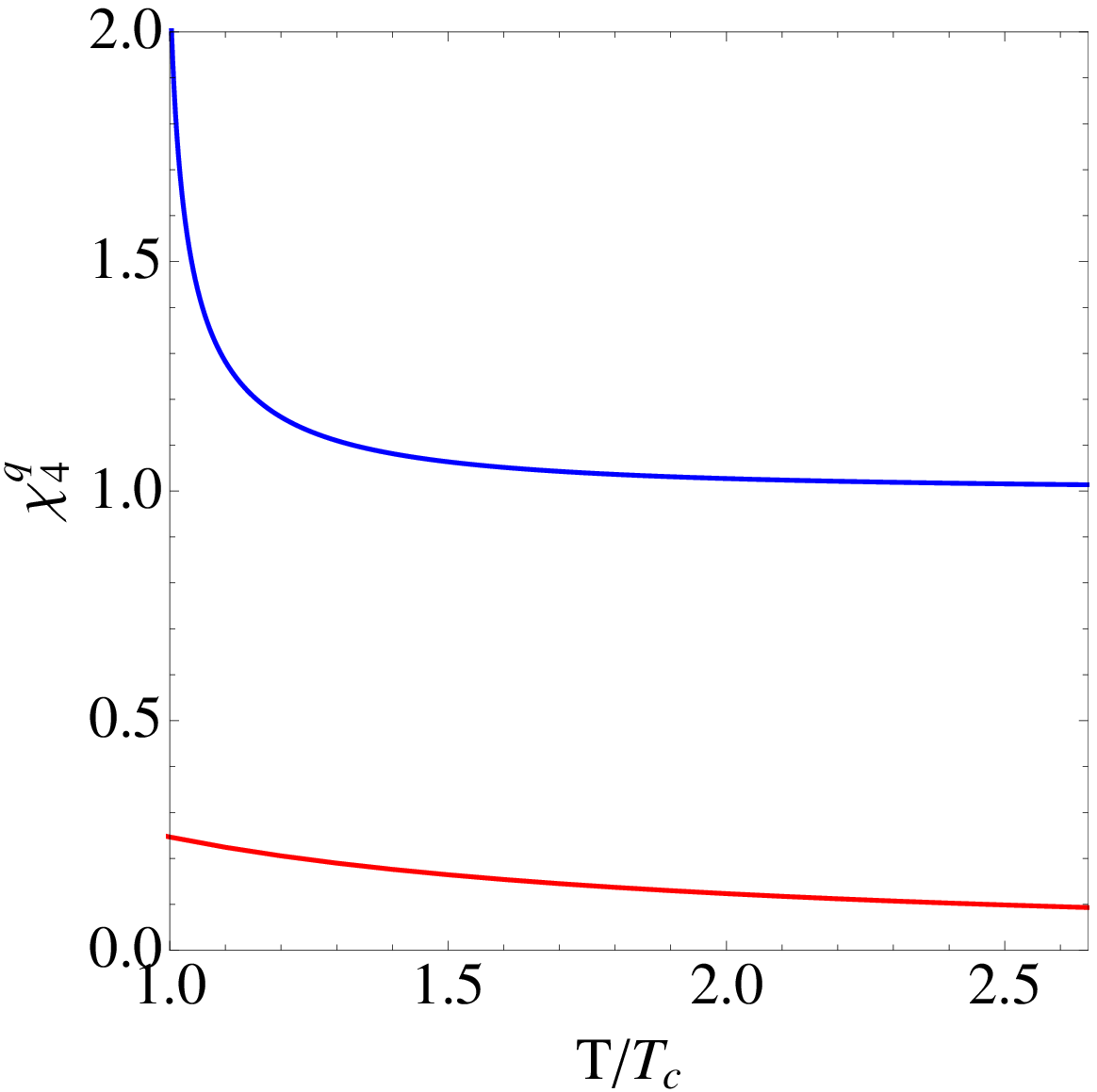} \; 
\includegraphics[width=0.3\textwidth]{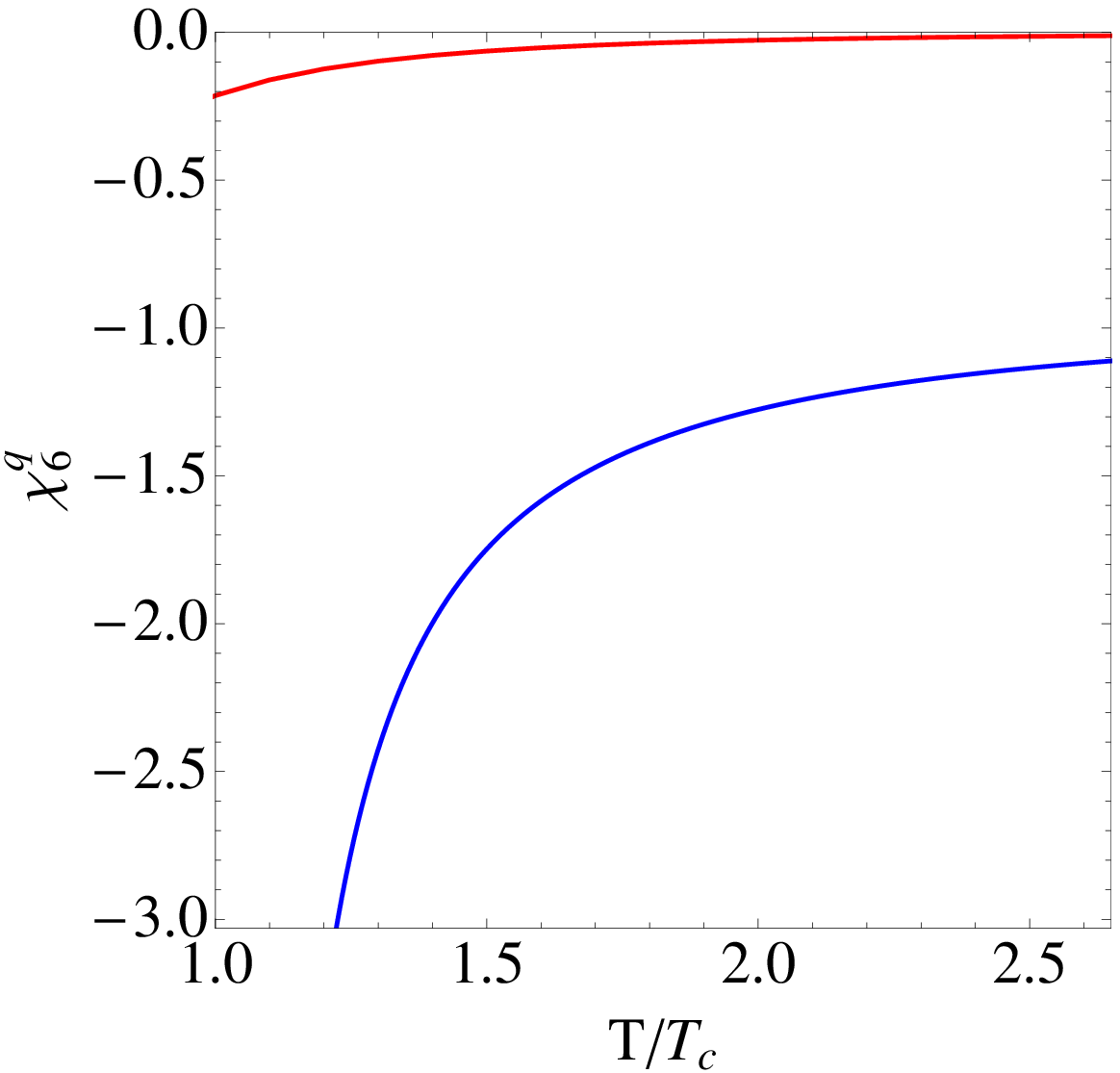}
\caption{Comparison of $\chi^q_{n=2,4,6}(T)$ from the D4/D8 model (red curves) and from the D3/D7 model (blue curves). Also shown for $\chi^q_2$ are lattice QCD results (black diamonds). }
\label{fig:comp}
\end{center}
\end{figure}

\subsection{Can quasi-particle models describe the conserved charge fluctuations?}

There have also been a lot of studies in the literature trying to understand these susceptibilities in terms of quasi-particle models \cite{Liao:2005pa,Ratti:2007jf,Sasaki:2006ww,Skokov:2010uh,Bluhm:2008sc,Dumitru:2012fw,Lin:2009ds}. Such models are based on the idea that the (complicated) effects from interactions in the many body setting may be approximately accounted for by a simple medium modification of single particle (the quasi-particle) properties, e.g. a temperature and density dependent mass. In the context of QGP, the quasi-particle models assume the plasma to be a gas of independent quasi-quarks and quasi-gluons whose masses change with temperature and density. At very high temperature, this may indeed be a very good approximation and the quasi-particle properties are computable from perturbation theories, as demonstrated e.g. in the hard thermal loop approach \cite{Braaten:1989mz,Blaizot:2001nr,Blaizot:2001vr}. In the regime $T\sim \Lambda_{QCD}$, however, there is no a priori reason for a simple quasi-particle picture to be a good approximation. The susceptibilities are sensitive to the properties of the conserved charge carriers (e.g. quasi-quarks), and one may expect them to provide a useful test  of the quasi-particle models. Indeed lattice results for the susceptibilities in the $1-2T_c$ region show rather distinctive patterns that can not be easily explained by e.g perturbative calculations. 

It is useful here to recall the early analysis in \cite{Liao:2005pa} regarding different possibilities of explaining the susceptibilities in this region. The authors of \cite{Liao:2005pa} found two scenarios that may describe the data at that time. The first is an unconstrained quasi-particle model with the quasi-quark mass strongly decreasing from $T_c$ toward high T which may arise from certain nonperturbative dynamics. (Note that the quasi-quark mass $M_q(T,\mu)$ bears derivatives $d^i M /d\mu^i$ that could provide additional contributions to susceptibilities and in fact within such a quasi-particle model one can generally establish a one-to-one correspondence between $\chi^q_2$ and $M(T,0)$, $\chi^q_4$ and $dM/d\mu(T,0)$, $\chi^q_6$ and $d^2M/d\mu^2(T,0)$, etc, as pointed out in \cite{Liao:2005pa}.)  This scenario however requires a rather light quasi-quark mass that is in contradiction with other lattice data as well as model results, and furthermore it can not be extended into the hadronic side. The second scenario is to include baryonic states from the hadronic gas to naturally continue into the region above $T_c$ with their masses increasing with $T$ (which mimics the gradual melting of these states). This latter scenario provides good description of prominent features in the data then (``peak'' in the 4-th order and ``wiggle'' in the 6-th order susceptibilities) and also naturally explains the higher order baryon-isospin correlations. It was therefore concluded in \cite{Liao:2005pa} that the susceptibilities data are in favor of the scenario with bound states above $T_c$. 

\begin{figure}[!h]
\begin{center}
\includegraphics[width=0.45\textwidth]{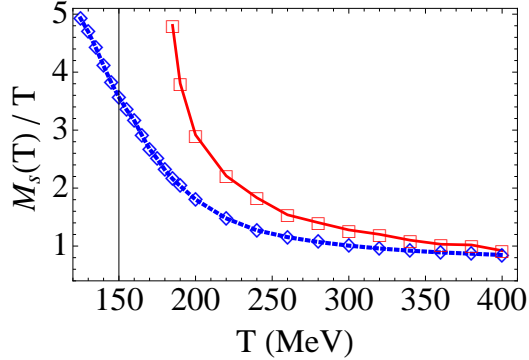} \;
\caption{Comparison of strange quasi-quark mass $M_s(T)/T$ extracted from $\chi^B_2$ \& $\chi^Q_B$ (red solid line with box symbols) with the same mass extracted from $\chi^S_2$ (blue dotted line with diamond symbols). }
\label{fig:Ms}
\end{center}
\end{figure}

Recently, high precision lattice data for susceptibilities of three conserved charges $B,Q,S$ (baryon number, electric charge, strangeness) in QGP became available \cite{Borsanyi:2011sw,Bazavov:2012jq}, and with these three quantities a much more stringent consistency test is in order for quasi-particle models which have only two adjustable quasi-quark masses for the light quarks and for the strange quark.  
Let us consider such a quasi-particle model with in-medium masses $M_l(T,\mu)$ for u,d quarks and $M_s(T,\mu)$ for s quark. To compute the fluctuations of $B,Q,S$ we start with their densities by summing over all quasi-quark contributions: 
\begin{eqnarray}
&& n_B =  2\cdot 3 \cdot  \sum_f \left(\frac{1}{3}\right) \cdot \int \frac{d^3p}{(2\pi)^3} \left[  \frac{1}{e^{(\sqrt{p^2 + M_f^2}-\mu_f)/T}+1} 
-  \frac{1}{e^{(\sqrt{p^2 + M_f^2} + \mu_f)/T}+1} \right] \\
&& n_Q =  2\cdot 3 \cdot  \sum_f \left(q_f \right) \cdot \int \frac{d^3p}{(2\pi)^3} \left[  \frac{1}{e^{(\sqrt{p^2 + M_f^2}-\mu_f)/T}+1} 
-  \frac{1}{e^{(\sqrt{p^2 + M_f^2} + \mu_f)/T}+1} \right]  \\
&& n_S =  2\cdot 3 \cdot   \left(-1\right) \cdot \int \frac{d^3p}{(2\pi)^3} \left[  \frac{1}{e^{(\sqrt{p^2 + M_s^2}-\mu_s)/T}+1} 
-  \frac{1}{e^{(\sqrt{p^2 + M_s^2} + \mu_s)/T}+1} \right] 
\end{eqnarray}
with $q_f$ is the quark electric charge with $q_u=2/3$ and $q_{d,s}=-1/3$. Note that the flavor-wise chemical potentials are related to the conserved charge chemical potentials via: $\mu_u=\mu_B/3 + 2\mu_Q/3$, $\mu_d=\mu_B/3 - \mu_Q/3$, $\mu_s=\mu_B/3 -  \mu_Q/3 - \mu_S$. The susceptibilities can then be computed via: 
\begin{eqnarray}
\chi^B_2(T) = \frac{\partial n_B}{\partial \mu_B} {\bigg |}_{all\; \mu's \to 0} \; , \; 
\chi^Q_2(T) = \frac{\partial n_Q}{\partial \mu_Q} {\bigg |}_{all\; \mu's \to 0} \; , \; 
\chi^S_2(T) = \frac{\partial n_S}{\partial \mu_S} {\bigg |}_{all\; \mu's \to 0}
\end{eqnarray}
Note that by charge conjugation symmetry the quasi-particle masses $M_f(T,\mu_f)$  is an even function of $\mu_f$ i.e. $dM/d\mu|_{\mu\to 0} \to 0$. Therefore one sees in the present approach that only $M_l(T)$ (u,d quasi-quark masses) and $M_s(T)$ (s quasi-quark mass) will appear in $\chi^B_2$ and $\chi^Q_2$ while the $M_s(T)$ alone will appear in $\chi^S_2$. This situation provides an opportunity to check the consistency of the quasi-particle model: one can extract the $M_l(T)$ and $M_s(T)$ from lattice data for $\chi^B_2$ and $\chi^Q_2$, and alternatively one can also extract the $M_s(T)$ from lattice data for $\chi^S_2$, and then confront the two ways of extracting the same quantity $M_s(T)$. Using the lattice data from \cite{Borsanyi:2011sw} we have done this exercise and the results are compared in Fig.\ref{fig:Ms}. While the two agree for temperatures greater than $\sim 2.5 T_c$,  one clearly sees a divergence of the two toward lower $T$ and the discrepancy becomes huge when $T\to T_c$. 

From this analysis using the state-of-the-art lattice QCD data, we conclude that simple quasi-particle models can not provide a consistent description of the conserved charge fluctuations in the $1-2T_c$ regime. There are of course varied ways of implementing quasi-particle models (e.g. starting from pressure rather than from conserved charge densities and including $T,\mu$ dependent Bag terms, adding Polyakov loops, etc) which may yield somewhat different results. But the main message is unlikely to change, namely there are very strong cross-flavor correlations (described by off-diagonal susceptibilities) in this regime that are generally missed by quasi-particle models.   

\subsection{Do off-diagonal susceptibilities imply bound states above $T_c$?}

Finally let us discuss the off-diagonal susceptibilities, e.g.  $\chi_{ud},\chi_{us}$ which provide direct information on the density-density correlations between particle species carrying different flavors. In quasi-particle models such cross-flavor correlations are missing, while in holographic models (and generically in large $N_c$ limit) such off-diagonal susceptibilities are $1/N_c$ suppressed as compared with the diagonal ones as recently shown by Casalderrey-Solana and Mateos in \cite{CasalderreySolana:2012np}. They are however very sensitive to the underlying degrees of freedom, in particular possible composite objects (such as mesonic states) that naturally carry multiple flavors. Since the QGP in the $1-2T_c$ regime is a strongly interacting system and the interactions between quarks/anti-quarks are still strong, it is plausible to have bound states (both meson and baryon like ones, and possibly colored bound states) above $T_c$ \cite{Shuryak:2004tx,Liao:2005hj}. The off-diagonal susceptibilities thus can be a good test of the existence of such bound states. In fact observables along this line were proposed quite some time ago: one example is the B-S correlation $C_{BS} \equiv - 3 <BS>/<S^2> $ \cite{Koch:2005vg} which is a sensitive probe of the strangeness carrying states; another example is the high-order B-I (I being the isospin) correlations $<B^2 I^2>/<B^2> <I^2>$ and $<B^4 I^2>/<B^4> <I^2>$ \cite{Liao:2005pa} which are particularly sensitive to existence of high baryon charge objects such as the survived baryonic bound states above $T_c$. Recent high precision lattice data e.g. for the $C_{BS}$ indeed show significant departure from  simple independent (quasi-)quark picture and the nice analysis in \cite{Ratti:2011au} appears to suggest that bound states are indispensable for explaining such off-diagonal susceptibilities. 

\begin{figure}[!h]
\begin{center}
\includegraphics[width=0.3\textwidth]{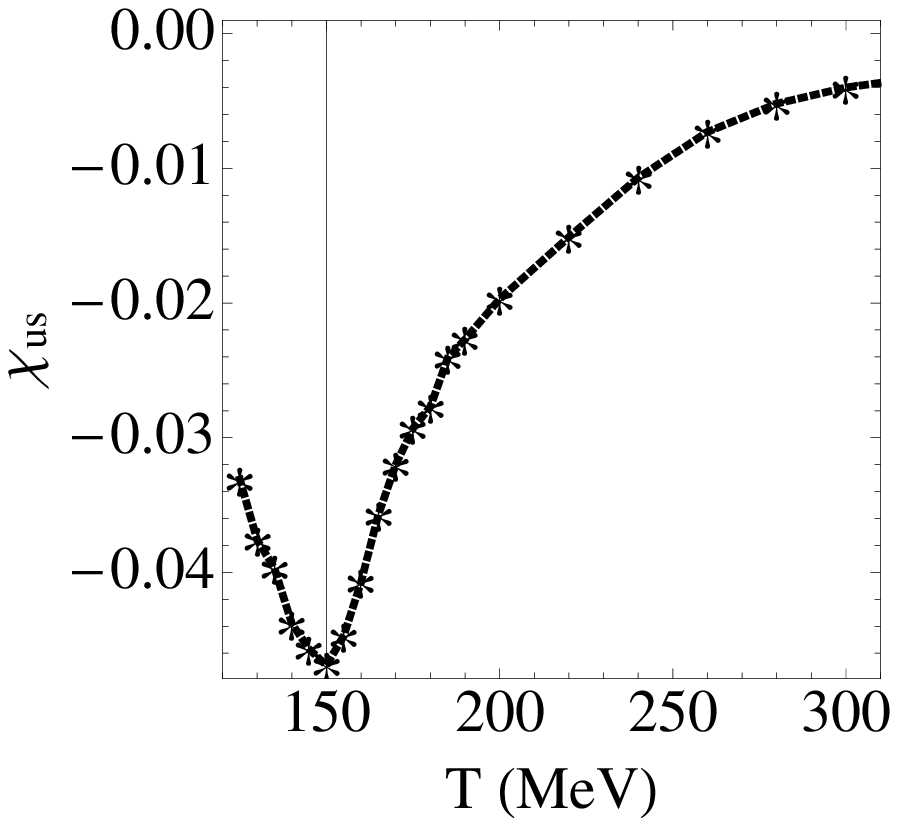} \; \; 
\includegraphics[width=0.3\textwidth]{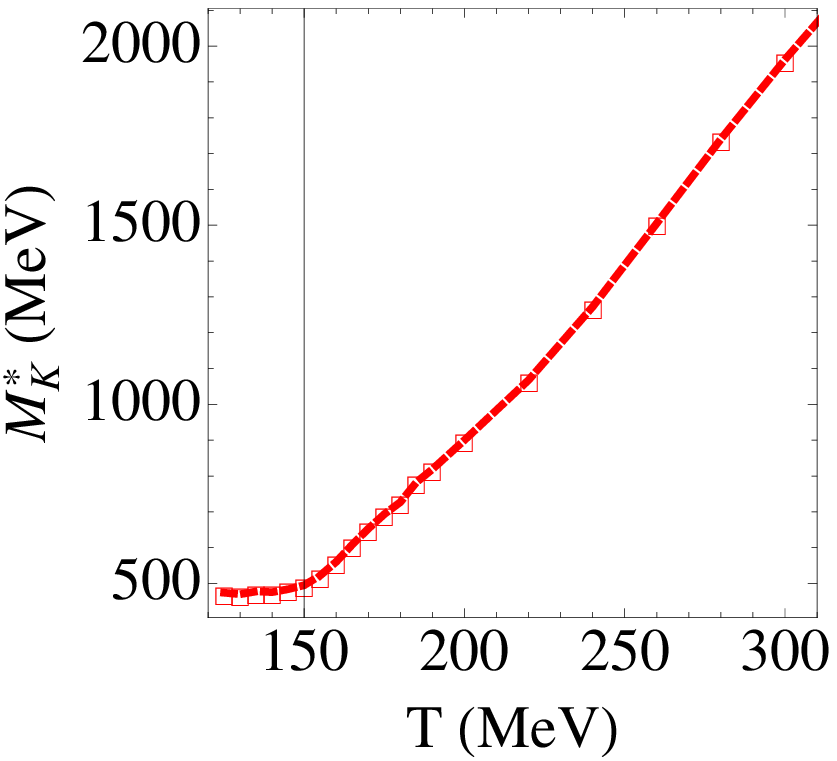} \; \;
\includegraphics[width=0.3\textwidth]{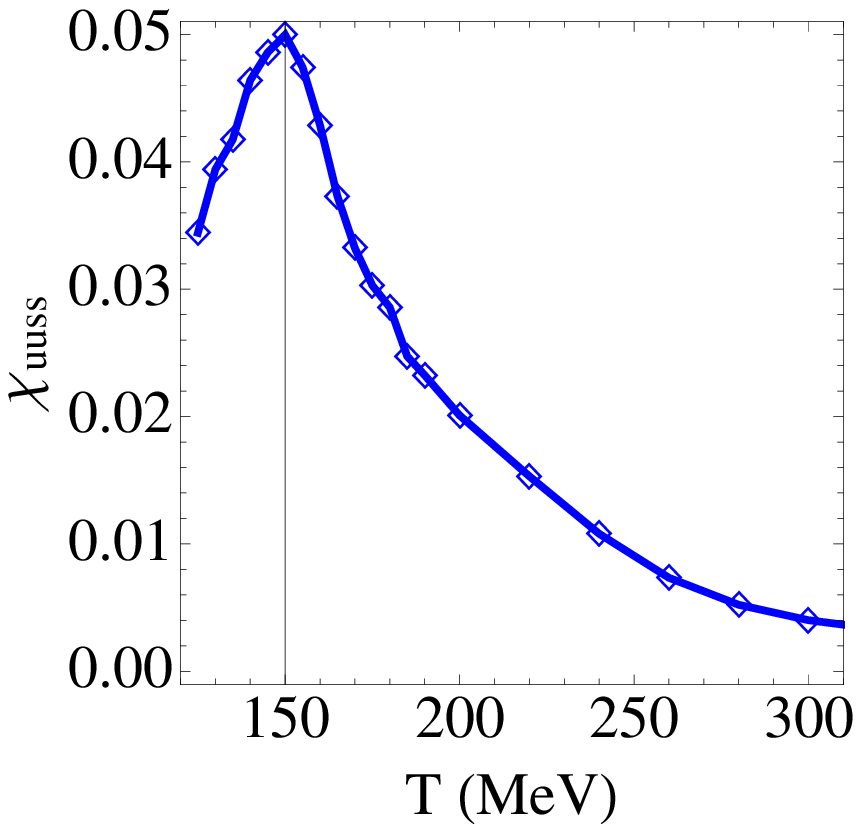} 
\caption{(left) lattice QCD results for $\chi_{us}$; (middle) extracted temperature dependent mass $M^*_K(T)$ for kaon states; (right) contributions of these kaon states to higher order off-diagonal susceptibilities $\chi_{uuss}$.}
\label{fig:chius}
\end{center}
\end{figure}

\subsubsection{Mesonic states}

Let us first focus on the lattice results for $\chi_{us}$ \cite{Borsanyi:2011sw} as shown in Fig.\ref{fig:chius} (left). Note that for a gas of quarks at high T one would expect $\chi_{us} \to 0$ while in the hadronic phase $T<T_c\approx 150\rm MeV$   the $\chi_{us}$ can be explained by a hadronic resonance gas model. There is strong negative contribution even above $T_c \approx 150MeV$  and continuing toward $\sim 2T_c$ --- note also that if one simply continues the vacuum hadrons' contribution to $T>T_c$ the curve would then continue going more and more negative rather than bending back. A possible explanation of this behavior could be the gradual melting of hadronic bound states with increasing temperature \cite{Liao:2005pa,Liao:2005hj,Ratti:2011au}.  
Among the hadronic states contributing to $\chi_{us}$, the two charged kaon states ($K^{\pm}$) are the lightest and most robust, and indeed they contribute negatively to $\chi_{us}$. For simplicity we assume  these states dominate the bound states' contribution to the $\chi_{us}$, and examine the implication of the lattice data. To mimic the ``melting'' of these states, we assume an effective mass $M_K^*(T)$ for these kaon states, and their contribution to the $\chi_{us}$ is then given by: 
\begin{eqnarray}
\chi_{us}^K =  - \frac{2}{2\pi^2} \int_0^\infty dx \frac{x^2 e^{\sqrt{x^2+(M_K^*/T)^2}}}{[e^{\sqrt{x^2+(M_K^*/T)^2}}-1]^2}
\end{eqnarray}
Then from the lattice data on $\chi_{us}$ one can extract the effective mass $M_K^*(T)$, as shown in the Fig.\ref{fig:chius} (middle). As one can see, the mass is essentially the physical vacuum value for $T<T_c$, while starts to rapidly increase for $T>T_c$ which means these kaon states are more and more strongly suppressed toward higher temperature. Of course such increasing mass is just an effective way of mimicking the melting of such states due to  increased screening of interaction potentials and in reality it may not be plausible to think of the kaon-like states with $\sim GeV$ mass. To see if such a way of mimicking the melting of bound states above $T_c$ is a good approximation, we can use the higher order off-diagonal susceptibilities to test it. Contributions of these kaon states to the $\chi_{uuss}$ (and $\chi_{uuss}=\chi_{usus}=-\chi_{usss}=-\chi_{uuus}$ from these states) based on the present assumption and extracted $M_K^*(T)$ are given in the Fig.\ref{fig:chius} (right), which can be tested by future lattice QCD data. We nevertheless point out that there are in principle contributions from various other mesons (e.g. excited charged kaon states) and baryons (e.g. $\Sigma,\Lambda,\Xi$ states) and some of the baryons' contributions may become more visible in higher order susceptibilities (as $\Sigma,\Xi$ carry more units of u or s flavors). On the other hand, if these higher mass, more loosely bound states could survive above $T_c$, there is no reason why the lowest kaon states wouldn't. 

\begin{figure}[!h]
\begin{center}
\includegraphics[width=0.55\textwidth]{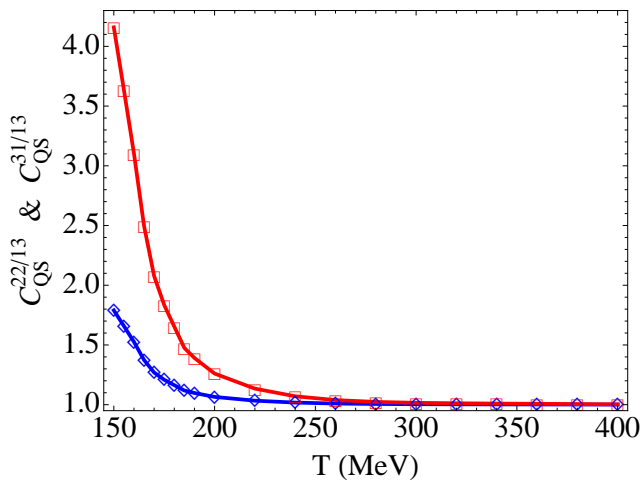} 
\caption{ The ratios $C_{QS}^{22/13}$ (blue diamond) and $C_{QS}^{31/13}$ (red box) for the 4th order baryon-electric charge correlations: see text for detailed discussions.}
\label{fig:CQS}
\end{center}
\end{figure}

Another place where the strange mesons (mainly kaons) can play an important role is the strangeness-electric charge correlations. On the quark side only strange quarks/anti-quarks contribution while on the hadronic side the kaons will contribution, and importantly the $K^{\pm}$ have their electric charges three times more than the quarks so they will be more and more prominent in higher and higher order S-Q correlations. For example, one can construct the following ratios of fourth-order susceptibilities 
\begin{eqnarray}
C_{QS}^{22/13}  \equiv 3\, \frac{<Q^2 S^2>}{<Q S^3 >} \;\; , \;\;  C_{QS}^{31/13}  \equiv 9\, \frac{<Q^3 S^1>}{<Q S^3 >}
\end{eqnarray}
They are normalized such that the strange quarks/anti-quarks will contribute unity. The $K^{\pm}$ mesons will contribute $3$ and $9$ respectively and will be more visible in the latter ratio. One can make a simple model to compute these ratios. If we assume at the second order susceptibilities in QGP, the $S^2$ is dominated by strange quarks while the $\chi_{us}$ is dominated by the $K^{\pm}$ mesons, then we can use the extracted $M_s(T)$ and $M^*_K(T)$ to make predictions for these two ratios: the results are shown in Fig.\ref{fig:CQS}. The patterns are quite clearly showing a gradual switch from meson-dominated contributions to the quark-dominated contributions from low to high temperature and the switching is slower in the $C_{QS}^{31/13} $ where the mesons should be more visible. At very quantitative level, the fraction from strange quarks in the plots may be a bit overestimated for two reasons: first  the $<S^2>$ should not be entirely due to strange quarks; second there are many other higher strange mesons and baryons states which will contribute. So the actual ratios may be slightly higher than the simple model estimations here, but we expect the patterns to be readily there and the numbers to be quite quantitative.

\subsubsection{Baryonic states}

\begin{figure}[!h]
\begin{center}
\includegraphics[width=0.31\textwidth]{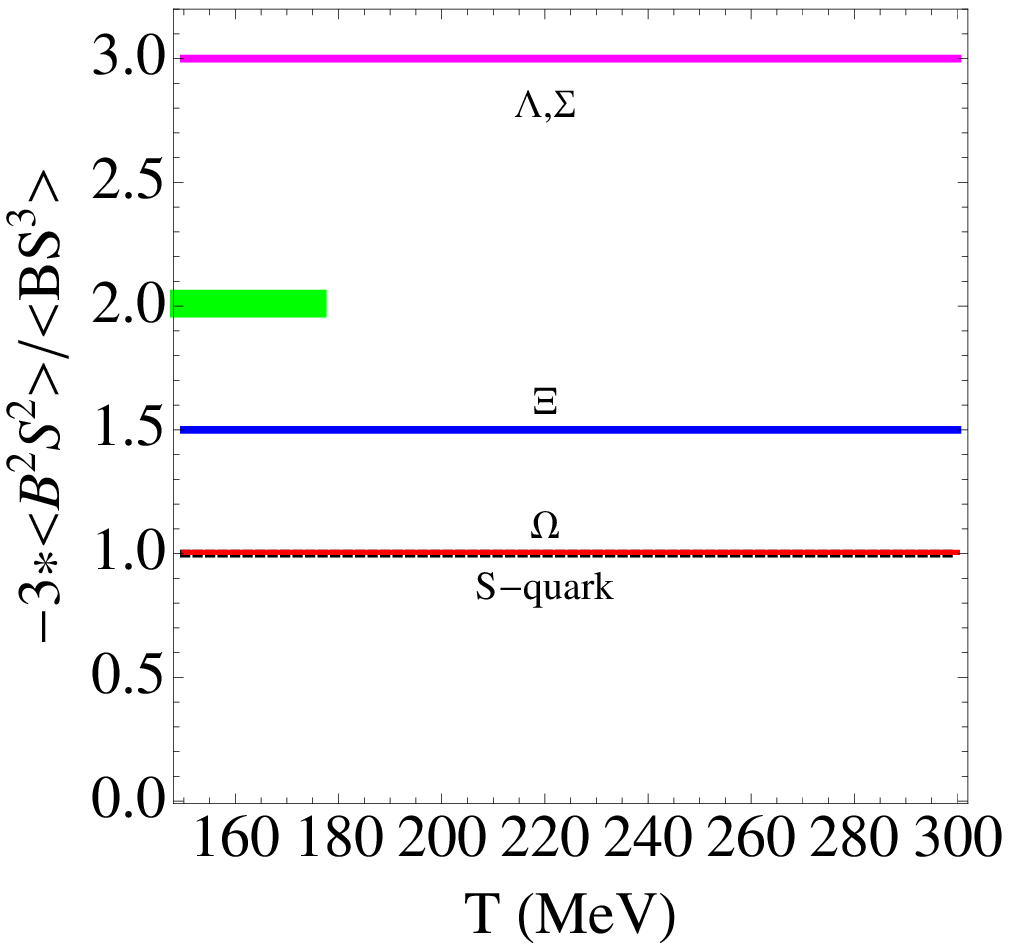} \; \; 
\includegraphics[width=0.3\textwidth]{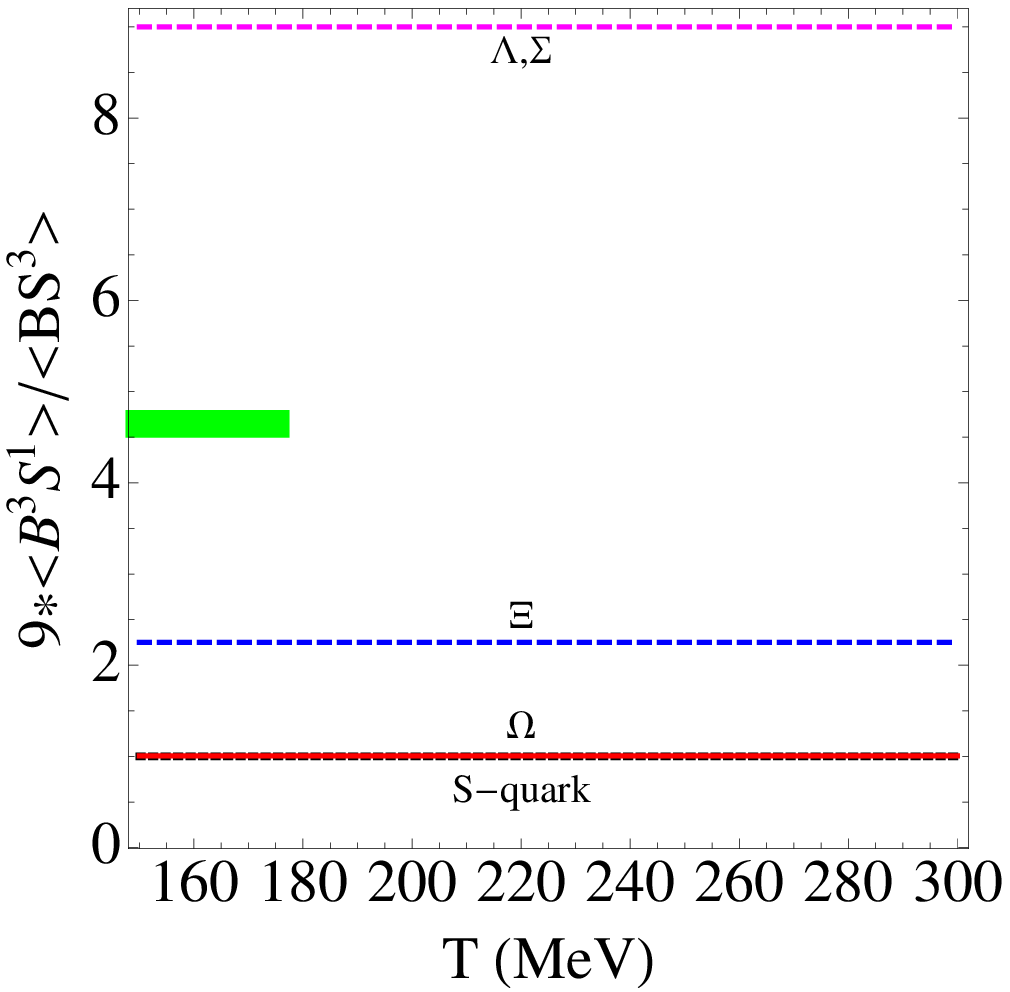} \; \; 
\includegraphics[width=0.3\textwidth]{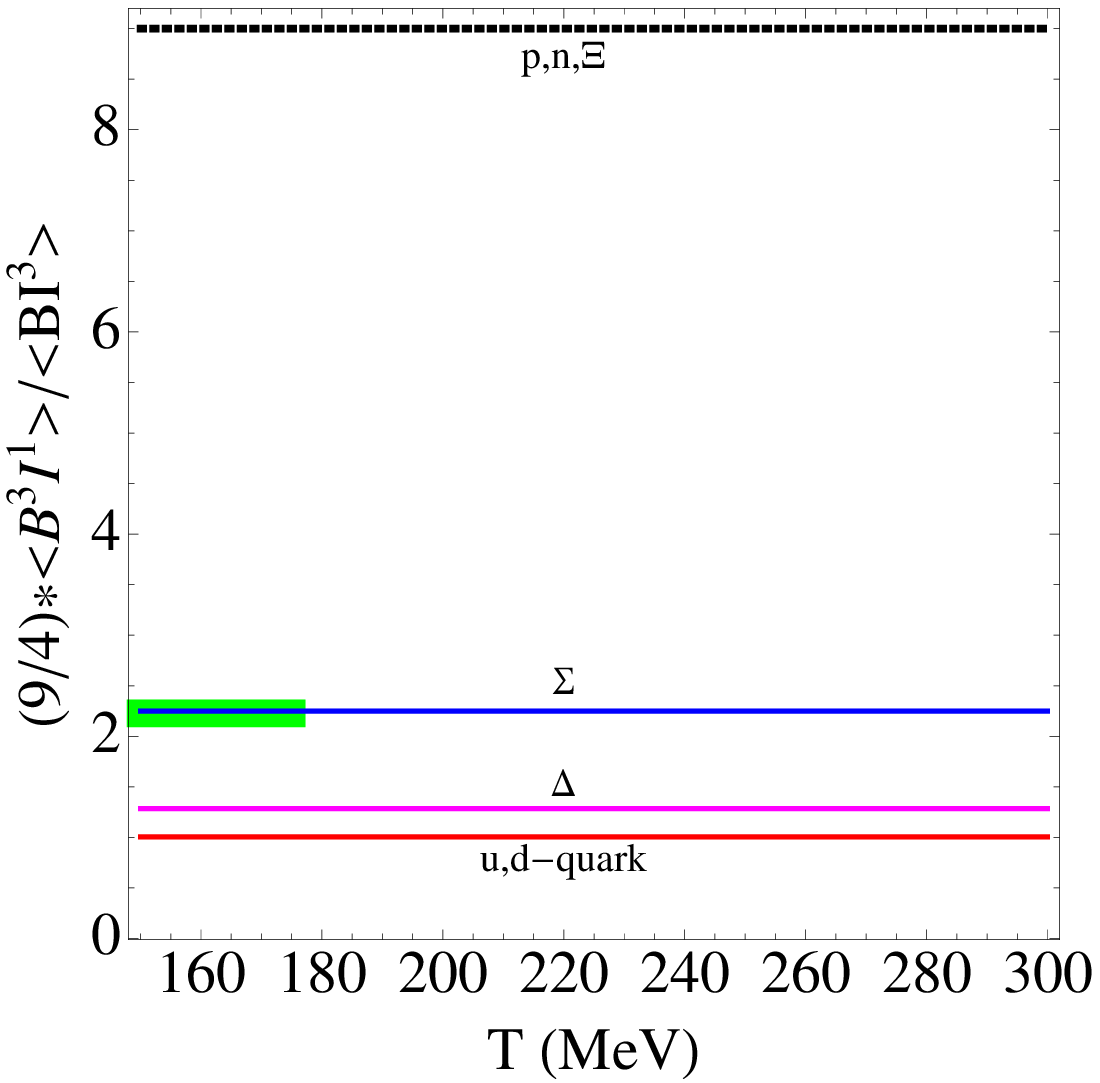} 
\caption{ Various baryonic and quark states' contributions to the proposed 4th order baryon-strangeness $C_{BS}^{22/13}$ (left), $C_{BS}^{31/13}$ (middle) and baryon-isospin correlations $C_{BI}^{31/13}$ (right), with the green bands indicating the value for the mixture of all barynoic states in Particle Data Book below 2GeV: see text for detailed discussions. }
\label{fig:BSBI}
\end{center}
\end{figure}

A more direct probe to ``tell'' if there is any baryonic bound state would be the direct correlations between baryon number and other conserved charges such as the baryon-isospin correlation proposed in \cite{Liao:2005pa} or the baryon-strangeness correlation proposed in \cite{Koch:2005vg}. Note however that in the case of baryon-strangeness $C_{BS}=-3<BS>/<S^2>$, there are still strange mesons' contributions to the $<S^2>$ in the denominator. (Similar issue exists for the baryon-isospin correlation as the normalization $<I^2>$ involves mesons' contributions.) In order to probe the purely baryonic states and avoid the complication with the mesons, one can use specific fourth order cumulants. For the baryon-strangeness correlations, one can study the following ratios: 
\begin{eqnarray}
C_{BS}^{22/13}  \equiv -3\, \frac{<B^2 S^2>}{<B S^3 >} \;\; , \;\;  C_{BS}^{31/13}  \equiv 9\, \frac{<B^3 S^1>}{<B S^3 >}
\end{eqnarray}
In these ratios, only strange baryons and strange quarks contribute to both the numerator and the denominator and they are normalized such that the pure strange quarks will give unity for both. In Fig.\ref{fig:BSBI} (left and middle) we indicate the values of these ratios for a single component made of various baryonic states based on their B and S quantum numbers. It is more difficult to make a quantitative estimate for these ratios, as the observables involve a lot of baryonic states with varied masses and medium modifications. We can nevertheless make a simple estimation for the region not too much above $T_c$ by a mixture of all baryonic states in Particle Data Book below 2GeV and the results are shown in the Fig.\ref{fig:BSBI} as the green bands.  If indeed the baryonic states survive above $T_c$ and gradually melts away, then these ratios should gradually switch in the $1-2T_c$ regime from the baryon-dominated values (as indicated by the lines for the baryonic states) to the quark-dominated values (the lowest line at unity). Furthermore the states with larger baryonic charges (the baryons) are enhanced more in the ratio $C_{BS}^{31/13}$ compared in the $C_{BS}^{22/13}$, so the approaching to the quark value with increasing temperature will be slower for $C_{BS}^{31/13}$.   All these patterns can be readily tested by the accurate lattice QCD data for 4th order off-diagonal susceptibilities.

Similarly for the baryon-isospin correlations, one can study the following ratio: 
\begin{eqnarray}
 C_{BI}^{31/13}  \equiv  \frac{9}{4} \, \frac{<B^3 I^1>}{<B I^3 >}
\end{eqnarray}
In the above ratio, only baryons with nonzero isospin and the u,d quarks contribute to both the numerator and the denominator and they are normalized such that the pure u,d quarks will give unity for both. In Fig.\ref{fig:BSBI} (right) we indicate the values of the ratio for a single component made of various baryonic states based on their B and S quantum numbers, and again we make a simple estimation for the region not too much above $T_c$ by a mixture of all baryonic states in Particle Data Book below 2GeV and the results are shown in the Fig.\ref{fig:BSBI} as the green band. Note for the isospin it is slightly more complicated and the $C_{BI}^{22/13}$ would be rather complex due to the opposite contributions from isospin partners (such as the protons and neutrons as well as the u and d quarks). We expect this ratio will gradually decrease from the green band and change toward the pure quark value when temperature is increased to $\sim 2T_c$.

Finally it is not difficult to generalize such ratios to the 6-th order susceptibilities that provide even more sensitive tests to the existence of bound states which future lattice data will be able to prove or disprove. 

\section{Summary}

In summary we have studied the conserved charge fluctuations, as quantified by the corresponding susceptibilities, in strongly interacting matter via a number of approaches. The holographic models motivated by QCD matter provide ways to overcome the challenge posed by the very strong interaction in many body setting, and we have used two types of such models, the D4/D8 and the D3/D7 models, to explore the patterns of conserved charge fluctuations in these systems. We have explicitly computed in both models the susceptibilities and certain combinations of them that are of physical interest,  with both analytical results when possible and numerical results when needed. While the susceptibilities from the two models show different behaviors in several aspects, they share the common feature that at very strong coupling higher order susceptibilities are suppressed and the conserved charge fluctuations become purely Guassian. This has not been known before, and it would be very interesting to see if this feature is universally true also in other very strongly coupled systems. 

In light of the state-of-the-art  lattice QCD results for (mostly second order) susceptibilities  with distinctive patterns in the $1-2T_c$ region, we have also examined the viability of different models aiming to explain these data. The holographic models show qualitatively interesting trends that are similar to lattice results but may not provide a quantitative explanation. The quasi-particle models are shown to contain inconsistency in explaining all conserved charge fluctuations due to the strong cross-flavor correlations that are missing in such models. It has also been demonstrated that hadronic bound states, which survive above $T_c$ and gradually melt away, provide a reasonable description of the off-diagonal susceptibilities. More future lattice QCD data for the higher order susceptibilities and particularly for the off-diagonal ones will be crucial for discriminating different models and for understanding the degrees of freedom in the strongly coupled quark-gluon plasma.

\section*{Acknowledgements}
The authors thank V. Koch, D. Mateos, L. McLerran, S. Mukherjee, K. Rajagopal, V. Skokov, M. Stephanov, and H. Yee for discussions and communications. JL is grateful to the RIKEN BNL Research Center for partial support. SS acknowledges support from the NSFC (Grant Nos. 10975084 and 11079024), RFDP (Grant No.20100002110080 ) and MOST (Grant No.2013CB922000).

\end{document}